\documentclass[11 pt,reqno]{amsart}

\usepackage{epsfig}
\usepackage{amsmath,amssymb}

\numberwithin{equation}{section}

\newcommand{\wt}{\widetilde}
\newcommand{\wh}{\widehat}

\newcommand{\R}{{\mathbb R}}
\newcommand{\C}{{\mathbb C}}

\newcommand{\Z}{{\mathbb Z}}

\newcommand{\Sq}{$\textrm{S}_q$}

\newcommand{\eq}{{\operatorname{eq}}}
\newcommand{\reg}{{\operatorname{reg}}}
\newcommand{\sing}{{\operatorname{sing}}}
\newcommand{\Gauss}{{\operatorname{Gauss}}}
\renewcommand{\phi}{\varphi}

\newcommand{\be}{\beta}
\newcommand{\ga}{\gamma}
\newcommand{\Ga}{\Gamma}

\newcommand{\ep}{\varepsilon}
\newcommand{\eps}{\epsilon}
\newcommand{\de}{\delta}

\newcommand{\om}{\omega}

\newcommand{\Ai}{{\,\rm Ai\,}}
\newtheorem{theo}{{\sc Theorem}}[section]
\newtheorem{cor}[theo]{{\sc Corollary}}
\newtheorem{lem}[theo]{{\sc Lemma}}
\newtheorem{prop}[theo]{{\sc Proposition}}
\newenvironment{example}{\medskip\noindent{\it Example:\/} }{\medskip}
\newenvironment{rem}{\medskip\noindent{\it Remark:\/} }{\medskip}

 2
 3
 4
 2

\begin{document}

\title[Asymptotics of the Partition Function]
{Asymptotics of the Partition Function of \\
a Random Matrix Model}

\author{Pavel Bleher}

\address{Department of Mathematical Sciences\\
               Indiana University-Purdue University Indianapolis\\
               402 N. Blackford Street\\
               Indianapolis, IN 46202, USA}

\email{bleher@math.iupui.edu.}

\author{Alexander Its}

\address{Department of Mathematical Sciences\\
               Indiana University-Purdue University Indianapolis\\
               402 N. Blackford Street\\
               Indianapolis, IN 46202, USA}

\email{itsa@math.iupui.edu}

\thanks{The first author was supported in part by the National Science 
  Foundation (NSF) Grants DMS-9970625 and  DMS-0354962. The second
  author was supported 
  in part by the NSF Grant DMS-0099812 and DMS-0401009.} 

\dedicatory{Dedicated to Pierre van Moerbeke on his sixtieth birthday.}

 \date{\today}

\begin{abstract} We prove a number of results concerning the
  large $N$ asymptotics of the free energy of a random matrix model
  with a polynomial potential $V(z)$. Our approach is based on a
  deformation $\tau_tV(z)$ of $V(z)$ to $z^2$, $0\le t<\infty$
  and on the use of the underlying integrable structures of the matrix model. 
  The main results   include (1) the existence of a full 
  asymptotic expansion in powers of $N^{-2}$ of the recurrence
  coefficients of the related orthogonal polynomials, for a one-cut 
  regular $V$; (2) the existence of a full 
  asymptotic expansion in powers of $N^{-2}$ of the free energy,
  for a $V$, which admits a one-cut regular deformation $\tau_tV$; (3) 
  the analyticity of the coefficients of the asymptotic expansions
  of the recurrence coefficients and the free energy,
  with respect to the coefficients of $V$; (4) the one-sided
  analyticity of the recurrent coefficients and the free energy 
  for a one-cut singular $V$; (5) the double scaling
  asymptotics of the free energy for a singular quartic polynomial
  $V$.      
\end{abstract}

\maketitle


\section{Introduction}\label{intro}

The central object of our analysis is the 
partition function of a random matrix model,
\begin{align}\label{f1}
Z_{N} &= \int_{-\infty}^{\infty}\ldots\int_{-\infty}^{\infty}
\prod_{1\leq j< k \leq N}(z_{j}-z_{k})^{2}
e^{-N\sum_{j=1}^{N}V(z_{j})}dz_{1}\ldots dz_{N} \nonumber\\
&= N!\prod_{n=0}^{N-1}h_{n},
\end{align}
where $V(z)$ is a polynomial,
\begin{equation}\label{f2}
V(z) = \sum_{j=1}^{2d}v_jz^{j}, \quad v_{2d} >0,
\end{equation}
and $h_{n}$ are the normalization constants of
the orthogonal polynomials on the line with respect
to the weight $e^{-NV(z)}$,
\begin{equation}\label{f3}
\int_{-\infty}^{\infty}P_{n}(z)P_{m}(z)e^{-NV(z)}dz
=h_{n}\delta_{nm}\,;\qquad
P_{n}(z) = z^{n} + \ldots
\end{equation}
In this work we are interested in the asymptotic expansion of the free
energy, 
\begin{equation}\label{f3a}
F_N=-\frac{1}{N^2}\ln Z_N,
\end{equation}
as $N\to\infty$. Our approach is based on the deformation $\tau_t$
of $V(z)$ to $z^2$,
\begin{equation}\label{f3b}
\tau_t:\; V(z)\to (1-t^{-1})z^2+V(t^{-1/2}z),\qquad 1\le t<\infty,
\end{equation}
so that
\begin{equation}\label{f3c}
\tau_1 V(z)=V(z),\qquad \tau_{\infty}V(z)=z^2,
\end{equation}
and our main reasults are the following:
\begin{enumerate}
\item\label{result1}{under the assumption that $V$ is one-cut regular
(for definitions see Section \ref{1cut_regular} below), we obtain a
full asymptotic expansion of the 
recurrence coefficients $\ga_n,\be_n$ of orthogonal polynomials in 
powers of $N^{-2}$, and we show the analyticity of the coefficients of
these asymptotic expansions  with respect to the
 coefficients $v_k$, $k=1,\dots, 2d$;}
\item\label{result2}{under the assumption that $\tau_t V$ is one-cut regular
 for $t\ge 1$, we prove the full asymptotic expansion of $F_N$ in
powers of $N^{-2}$, and we show the analyticity of the coefficients of
 the asymptotic expansion
with respect to  $v_k$, $k=1,\dots, 2d$;}
\item\label{result3}{under the assumptions that (i) $V$ is singular,
(ii) the equilibrium measure of $V$ is nondegenerate at the end-points, and
(iii) $\tau_t V$ is one-cut regular
for  $t>1$, we prove that the coefficients of
 the asymptotic expansion of the free energy of $\tau_tV$  for $t>1$
can be analytically continued to $t=1$; }
\item\label{result4}{for the singular quartic polynomial,
  $V(z)=(z^4/4)-z^2$, we obtain the double scaling asymptotics of the
  free energy of $\tau_t V(z)$ where $t-1$ is of the order
of $N^{-2/3}$; we prove that this asymptotics is a sum of a
  regular term, coming as a limit of the asymptotic expansion for
  $t>1$, and a singular term, which has the form of the logarithm of
  the Tracy-Widom distribution function.}
\end{enumerate}
In result (\ref{result2}),
the existence of a full asymptotic expansion of the free energy
in powers of $N^{-2}$ was first proved
by Ercolani and McLaughlin \cite{EM}, under the assumption
that the coefficients of 
$V$ are small. It was used in \cite{EM} to make rigorous the
Bessis-Itzykson-Zuber topological 
expansion \cite{BIZ} related to counting Feynman graphs on Riemannian surfaces.
The approach of Ercolani and McLaughlin is based on an asymptotic
analysis of the solution of the Riemann-Hilbert problem, and it is
very different from our approach, which is based on the deformation
equations. Also in result (\ref{result2}),  
our proof of the analyticity of the coefficients of the asymptotic
  expansion with respect to $v_k$
uses an important result of Kuijlaars and McLaughlin \cite{KM}, that
the Jacobian of the map of the end-points of the equlibrium measure
to the basic set of integrals (see Section \ref{1cut_regular} below)
is nonzero. In result (\ref{result3}),
the existence of an analytic continuation of the free energy to the
critical point (the one-sided analyticity) from the one-cut side
 was proved by Bleher and Eynard  for a
nonsymmetric singular quartic polynomial, see
the paper \cite{BE}, where, in fact, the one-sided
analyticity was proved from the both sides, one-cut and two-cut, and 
a phase transition of the third order was shown. Thus, 
result (\ref{result3})
gives an extension of the result of \cite {BE} to a general singular 
$V$ from the one-cut side. Observe that the analytic behavior of the
free energy from the multi-cut side can be different for different
singular $V$ and it requires a special investigation.
  In result (\ref{result4}), to derive and to prove the double
scaling asymptotics of the free energy we use and slightly extend 
the double scaling
asymptotics of the recurrent coefficients, obtained in our paper
\cite{BI2}. In addition, we develop the Riemann-Hilbert approach of
\cite{DKMVZ} for the case when $t=1+cN^{-(2/3)+\eps}$, where
$c,\eps>0$. In this case the lenses thickness vanishes as
$N^{-1/3}$ but this is enough to estimate the jump on the
lenses by $e^{-CN^{\eps}}$ and to apply the methods of \cite{DKMVZ}. 

The set up of the rest of the paper is the following.  In Section
\ref{start} we derive formulas which describe the deformation of the
recurrence coefficients and the free energy for a finite $N$, 
under deformations of $V$. Here, we make use of the integrability
of the matrix model, and we refer the reader to  excellent recent surveys
of van Moerbeke \cite{vMo1}, \cite{vMo2} on different modern aspects as well
as the history of the matter. 
In Section \ref{finite_N} we use the deformation  $\tau_tV(z)$ 
to obtain an integral representation of the free energy
for a finite $N$. In Section \ref{1cut_regular} we obtain different
results concerning the analyticity of the equilibrium measure for
the $q$-cut regular case. In Section \ref{rec_coeff} we obtain one of
  our main results about the asymptotic expansion of recurrence
  coefficients in the one-cut regular case.  This is applied then in
  Section \ref{free_energy} to obtain the asymptotic expansion of the
  free energy, assuming that $\tau_t V$ is one-cut regular for
  $t\in[1,\infty)$. In Section \ref{exact_formula} we derive an exact
  formula for the limiting free energy in the case when $V$ is
  an even one-cut regular polynomial.
  In Section \ref{one-sided} we obtain a number of results
  concerning the one-sided analyticity for singular $V$. Finally,
  in Section \ref{DSL} we obtain the double scaling asymptotics of the free
  energy for the singular quartic polynomial $V$.

\section{Deformation Equations for Recurrence Coefficients and 
Partition Function}\label{start}

Define the psi-functions as
\begin{equation}\label{f4}
\psi_n(z)=\frac{1}{\sqrt{h_n}}P_n(z)e^{-\frac{V(z)}{2}}.
\end{equation}
Then 
\begin{equation}\label{f5}
\int_{-\infty}^{\infty}\psi_{n}(z)\psi_{m}(z)\,dz
=\delta_{nm}\,.
\end{equation}
The psi-functions satisfy the three term recurrence relation,
\begin{equation}\label{f6}
z\psi_{n}(z)
=\ga_{n+1}\psi_{n+1}(z) + \be_n\psi_n(z)+ \ga_{n}\psi_{n-1}(z),
\end{equation}
where
\begin{equation}\label{f7}
\ga_n=\sqrt{\frac{h_n}{h_{n-1}}}.
\end{equation}
Set
\begin{equation}\label{g1}
\vec\Psi(z)
=\begin{pmatrix}
\psi_0(z) \\
\psi_1(z) \\
\psi_2(z) \\
\vdots 
\end{pmatrix}
\end{equation}
Then (\ref{f6}) can be written in the matrix form as
\begin{equation}\label{g2}
z\vec\Psi(z)=Q\vec\Psi(z),\quad
Q=
\begin{pmatrix}
\be_0 & \ga _1 & 0 & 0 & 0 & \dots \\
\ga_1 & \be_1 & \ga _2 & 0 & 0 & \dots  \\
0 & \ga_2 & \be_2 & \ga _3 & 0 & \dots  \\
0 & 0 & \ga_3 & \be_3 & \ga_4 & \dots  \\
0 & 0 & 0 & \ga_4 & \be_4 & \dots  \\
\vdots & \vdots & \vdots & \vdots & \vdots & \ddots
\end{pmatrix}
\end{equation}

Observe that $Z_N$, $h_n$, $\ga_n$, $\be_n$ are functions 
of the coefficients $v_1,\ldots,v_{2d}$ of the polynomial
$V(z)$. We will be interested in exact expressions for the derivatives
of $Z_N$, $h_n$, $\ga_n$, $\be_n$ with respect to $v_k$.
Set
\begin{equation}\label{g3} 
\tilde v_k=Nv_k,\qquad k=1,\dots,2d.
\end{equation}

\begin{prop}\label{prop1} We have the following relations: 
\begin{align}\label{prop11}
\frac{\partial \ln h_{n}}{\partial \tilde v_k}
&= -[Q^k]_{nn},\\
\label{prop12}
\frac{\partial\ga_{n}}{\partial \tilde v_k}
&= \frac{\ga_n}{2}\left([Q^k]_{n-1,n-1}-[Q^k]_{nn}\right),\\
\label{prop13}
\frac{\partial\be_{n}}{\partial \tilde v_k}
&= \ga_n[Q^k]_{n,n-1}-\ga_{n+1}[Q^k]_{n+1,n},
\end{align}
where $[Q^k]_{nm}$ denotes the $nm$-th element of the matrix $Q^k$;
$n,m=0,1,2,\ldots$.
\end{prop}

\begin{proof} Formula (\ref{prop11}) is proven in \cite{Eyn}.
By (\ref{f7}), it implies (\ref{prop12}). Let us prove
(\ref{prop13}). Introduce the vector function
\begin{equation}\label{prop14} 
\vec\Psi_n(z)=
\begin{pmatrix}
\psi_{n-1}(z) \\
\psi_n(z)
\end{pmatrix}.
\end{equation}
As shown in \cite{Eyn} (see also \cite{BEH}), it satisfies the deformation equation
\begin{equation}\label{prop15} 
\frac{\partial\vec\Psi_n}{\partial \tilde v_k}=U_k(z;n)\vec\Psi_n(z),
\end{equation}
where
\begin{equation}\label{prop16} 
\begin{aligned}
U_k(z;n)&=\frac{1}{2}
\begin{pmatrix}
z^k-[Q^k]_{n-1,n-1} & 0 \\
0 & [Q^k]_{nn}-z^k
\end{pmatrix}\\
{}&+\ga_n
\begin{pmatrix}
{\rm }[Q(z;k-1)]_{n,n-1}  & -[Q(z;k-1)]_{n-1,n-1} \\ 
{\rm }[Q(z;k-1)]_{nn}  & -[Q(z;k-1)]_{n,n-1} 
\end{pmatrix},
\end{aligned}
\end{equation}
and 
\begin{equation}\label{prop17} 
Q(z;k-1)=\sum_{j=0}^{k-1} z^jQ^{k-1-j}.
\end{equation}
From (\ref{f6}),
\begin{equation}\label{prop18} 
\vec\Psi_{n+1}=\frac{1}{\ga_{n+1}}U(z;n)\vec\Psi_n(z),
\quad U(z;n)=
\begin{pmatrix}
0 & \ga_{n+1} \\
-\ga_n & z-\be_n
\end{pmatrix}.
\end{equation}
The compatibility condition of (\ref{prop16}) and (\ref{prop18}) is
\begin{equation}\label{prop19} 
\frac{\partial U(z;n)}{\partial \tilde v_k}=U_k(z;n+1)U(z;n)-U(z;n)U_k(z;n)
+\frac{1}{\ga_{n+1}}\frac{\partial\ga_{n+1}}{\partial \tilde v_k}
U(z;n).
\end{equation}
By restricting this equation to the element 22, we obtain
(\ref{prop13}). Proposition \ref{prop1} is proved.
\end{proof}

We will be especially interested in the derivatives with respect to $\tilde v_2$.
For $k=2$, Proposition \ref{prop1} gives that
\begin{align}\label{k1}
\frac{\partial\ln h_{n}}{\partial \tilde v_2}
&= -\ga_n^2-\be_n^2-\ga_{n+1}^2,\\
\label{k2}
\frac{\partial\ga_{n}}{\partial \tilde v_2}
&= \frac{\ga_n}{2}\left(\ga_{n-1}^2+\be_{n-1}^2-\ga_{n+1}^2-\be_n^2\right),\\
\label{k3}
\frac{\partial\be_{n}}{\partial \tilde v_2}
&= \ga_n^2\be_{n-1}+\ga_n^2\be_n
-\ga_{n+1}^2\be_n-\ga_{n+1}^2\be_{n+1}.
\end{align}
Observe that all these expressions are local in $n$,
so that they depend only on the recurrent coefficients with
indices which differ from $n$ by a fixed number.
Our next step will be to get a local expression for
the second derivative of $Z_n$.

\begin{prop}\label{prop2} We have the following relation: 
\begin{equation}\label{prop21}
\frac{\partial^2\ln Z_N}{\partial \tilde v_2^2}
= \ga_N^2\left(\ga_{N-1}^2+\ga_{N+1}^2
+\be_N^2+2\be_N\be_{N-1}+\be_{N-1}^2\right).
\end{equation}
\end{prop}

\begin{proof} For the sake of brevity we denote
  $(')=\frac{\partial}{\partial \tilde v_2}$. 
From (\ref{k1})-(\ref{k3}) we obtain that
\begin{equation}\label{prop22}
\begin{aligned}
(\ln h_n)''&=-2\ga_n\ga_n'-2\be_n\be_n'-2\ga_{n+1}\ga_{n+1}'
=-\ga_n^2\left(\ga_{n-1}^2+\be_{n-1}^2-\ga_{n+1}^2-\be_n^2\right)\\
&-2\be_n\left(\ga_n^2\be_{n-1}+\ga_n^2\be_n
-\ga_{n+1}^2\be_n-\ga_{n+1}^2\be_{n+1}\right)
-\ga_{n+1}^2\left(\ga_n^2+\be_n^2-\ga_{n+2}^2-\be_{n+1}^2\right)\\
&=I_{n+1}-I_n,
\end{aligned}
\end{equation}
where
\begin{equation}
I_n=\ga_n^2\left(\ga_{n-1}^2+\ga_{n+1}^2
+\be_n^2+2\be_n\be_{n-1}+\be_{n-1}^2\right).
\end{equation}
From (\ref{f1}) and (\ref{prop22}) we obtain now the telescopic sum,
\begin{equation}\label{prop23}
(\ln Z_N)''=
\sum_{n=0}^{N-1} (\ln h_n)''=\sum_{n=0}^{N-1}(I_{n+1}-I_n)
=I_N-I_0.
\end{equation}
Observe that $I_0=0$, because $\ga_0=0$, hence (\ref{prop21}) follows.
\end{proof}

\begin{rem} 
When $k=1$, Proposition \ref{prop1} gives that
\begin{align}\label{k1a}
\frac{\partial\ln h_{n}}{\partial \tilde v_1}=-\be_n,
\quad
\frac{\partial\ga_{n}}{\partial \tilde v_1}
= \frac{\ga_n}{2}\left(\be_{n-1}-\be_n\right),
\quad
\frac{\partial\be_{n}}{\partial \tilde v_1}
= \ga_n^2-\ga_{n+1}^2,
\end{align}
hence
\begin{equation}\label{prop21a}
\frac{\partial^2\ln Z_N}{\partial \tilde v_1^2}=\ga_N^2.
\end{equation}
Similar formulae can be derived also for $k\ge 3$,
but they become complicated.
\end{rem}

{\it Remark:} 
For the case of even potentials, equations (\ref{prop14} - \ref{prop19}),
as well as the statement of Proposition \ref{prop1} were
obtained in \cite{FIK}. It also worth noticing that in the even
case, differential-difference equation (\ref{prop12}) is the well-known
Volterra hierarchy whose integrability was
first established in 1974 - 75 in the pioneering works of Flaschka
\cite{F}, Kac and van Moerbeke \cite{KvM},
and Manakov \cite{Man}, and whose particular case (\ref{k2}) is the 
classical Kac-van Moerbeke discrete version of the KdV equation \cite{KvM}.

{\it Remark:}  
For the case of the even quartic potential $V(z) = v_{2}z^2 + v_{4}z^4$,
Proposition \ref{prop2} was proven in \cite{FIK0}.

\section{Free Energy for a Finite $N$} \label{finite_N}

In terms of $v_2$ formula (\ref{prop21}) reduces to the following:
\begin{equation}\label{sa3}
\frac{\partial^{2}F_N}{\partial v^2_2}
=-\ga_N^2\left(\ga_{N-1}^2+\ga_{N+1}^2
+\be_N^2+2\be_N\be_{N-1}+\be_{N-1}^2\right)\,,
\end{equation}
where $F_N$ is the free energy, see (\ref{f3a}).

The main problem we will be interested in is an asymptotics 
of the free energy as $N\to\infty$. Our approach will
be based on a deformation of the polynomial $V(z)$ to the quadratic
polynomial $z^2$. To that end we set
\begin{equation}\label{sa4a}
W(z)=V(z)-z^2,
\end{equation}
and we define a one-parameter family of polynomials, 
\begin{equation}\label{sa5}
V(z;t) = z^2+W\left(\frac{z}{\sqrt t}\right); 
\qquad t\ge 1\,.
\end{equation}
Then obviously,
\[
V(z;1)=V(z),\qquad V(z;\infty)=z^2.
\]
It is convenient to introduce the operator $\tau_t$, see (\ref{f3b}).
Then $V(z;t)=\tau_t V(z)$.
The operators $\tau_t$ satisfy the group property.

\begin{prop} \label{group}
\begin{equation}\label{group1}
\tau_t\tau_s=\tau_{ts}.
\end{equation}
\end{prop}

\begin{proof} We have that
\begin{equation}\label{group2}
\begin{aligned}
\tau_t(\tau_s(V(z)))&=\tau_t((1-s^{-1})z^2+V(s^{-1/2}z))
=(1-t^{-1}z^2)+(1-s^{-1})t^{-1}z^2+V(t^{-1/2}s^{-1/2}z)\\
&=(1-t^{-1}s^{-1})z^2+V(t^{-1/2}s^{-1/2}z)
=\tau_{ts}(V(z)).
\end{aligned}
\end{equation}
Proposition \ref{group} is proved.
\end{proof}
 
Let $Z_N=Z_N(t)$ be the partition
function (\ref{f1})
for the polynomial $V(z;t)$ and $F_N=F_N(t)$ the
corresponding free energy.

\begin{prop}\label{diff_t}
\begin{equation}\label{sa7}
\frac{\partial^{2}F_N(t)}{\partial t^2}
=-\frac{1}{t^2}\left\{\ga_N^2(t)\left[\ga_{N-1}^2(t)+\ga_{N+1}^2(t)
+\be_N^2(t)+2\be_N(t)\be_{N-1}(t)+\be_{N-1}^2(t)\right]-\frac{1}{2}\right\}, 
\end{equation}
where $\ga_n(t),\be_n(t)$ are the recurrence coefficients 
of orthogonal polynomials with respect to the weight $e^{-NV(z;t)}$.
\end{prop}

\begin{proof}
By the change of variables $z_j=\sqrt t\,u_j$, we obtain from
(\ref{f1}) that
\begin{equation}\label{sa8}
Z_{N}(t) =t^{N^2/2} \hat Z_N(t)\,,
\end{equation}
where
\begin{equation}\label{sa9}
\hat Z_N(t)=\int_{-\infty}^{\infty}\ldots\int_{-\infty}^{\infty}
\prod_{1\leq j< k \leq N}(u_{j}-u_{k})^{2}
e^{-\sum_{j=1}^{N}N\hat{V}(u_{j};t)}du_{1}\ldots du_{N}
\end{equation}
is the partition function for 
\begin{equation}\label{sa10}
\hat V(u;t)=V(\sqrt t\,u;t)=tu^2+W(u)\,.
\end{equation}
Hence
\begin{equation}\label{sa12}
\hat F_N(t)\equiv -\frac{1}{N^2}\,\ln \hat Z_N(t)
=-\frac{1}{N^2}\,\ln [-t^{N^2/2} Z_N(t)]=F_N(t)+\frac{\ln t}{2}\,.
\end{equation}
By (\ref{sa3}),
\begin{equation}\label{sa11}
\frac{\partial^{2}\hat F_N(t)}{\partial t^2}
=-\hat\ga_N^2(t)\left(\hat\ga_{N-1}^2(t)+\hat\ga_{N+1}^2(t)
+\hat\be_N^2(t)+2\hat\be_N(t)\hat\be_{N-1}(t)+\hat\be_{N-1}^2(t)\right)\,,
\end{equation}
where $\hat \ga_n(t)$, $\hat\be_n(t)$ are the recurrence coefficients 
of the  orthogonal polynomials 
\begin{equation}\label{sa13}
\hat P_n(u;t)=t^{-n/2}P_n(\sqrt t\,u;t)
\end{equation}
with respect to the weight $e^{-N\hat V(u;t)}$.
Since
\begin{equation}\label{sa13a}
(\sqrt t\,u)\,P_n(\sqrt t\,u;t)=P_{n+1}(\sqrt
t\,u;t)+\be_n(t)P_n(\sqrt t\,u;t)+\ga_n^2(t)P_{n-1}(\sqrt t\,u;t)\,, 
\end{equation}
we obtain that
\begin{equation}\label{sa14}
\hat \ga_n(t)=t^{-1/2}\ga_n(t)\,,\quad \hat \be_n(t)=t^{-1/2}\be_n(t)\,,
\end{equation}
hence from (\ref{sa11}), (\ref{sa12}) we obtain that
\begin{equation}\label{sa15}
\frac{\partial^{2}}{\partial t^2}\left( F_N(t)+\frac{\ln t}{2}\right)
=-\frac{\ga_N^2(t)}{t^2}\left[\ga_{N-1}^2(t)+\ga_{N+1}^2(t)
+\be_N^2(t)+2\be_N(t)\be_{N-1}(t)+\be_{N-1}^2(t)\right]\,,
\end{equation}
which implies (\ref{sa7}).
Proposition \ref{diff_t} is proven.
\end{proof}

We would like to integrate formula (\ref{sa7}). To that end we
need an asymptotic behavior of the recurrence coefficients 
$\ga_n,\be_n$, $n=N-1,\,
N,\,N+1$ as $t\to\infty$. For a  finite $N$ it is easy.

\begin{prop}\label{finite_n} Assume that $n$ and $N$ are fixed.
Then, as  $\,t\to\infty$,
\begin{equation}\label{fin1}
\ga_n(t)=\sqrt{\frac{n}{2N}}+ O(t^{-1/2}),\quad
\be_n=O(t^{-1/2})\,.
\end{equation}
\end{prop}

\begin{proof} When $t=\infty$, $V(z;t)=z^2$, hence $\ga_n(\infty)$,
$\be_n(\infty)$ 
are recurrence coefficients for the Hermite polynomials,
\begin{equation}\label{fin2}
\ga_n(\infty)=\sqrt{\frac{n}{2N}}\,,\quad
\be_n(\infty)=0.
\end{equation}
When $t$ is finite, the orthogonal polynomials and 
recurrence coefficients can be obtained through the Gramm-Schmidt 
orthogonalization algorithm. Since for any moment we have the 
relation,
\begin{equation}\label{fin3}
\int_{-\infty}^\infty z^ke^{-N\left(z^2+W(z/\sqrt{t})\right)}dz=
\int_{-\infty}^\infty z^ke^{-Nz^2}dz+O(t^{-1/2})\,,
\end{equation}
(\ref{fin1}) follows.
\end{proof}

From Propositions  \ref{diff_t} and \ref{finite_n} we obtain
the following formula for $F_N$.

\begin{theo}\label{F_N}
\begin{equation}\label{fin4}
\begin{aligned}
F_N(t)=F_N^{\Gauss}+\int_t^\infty \frac{t-\tau}{\tau^2}
&\left\{\ga_N^2(\tau)\left[\ga_{N-1}^2(\tau)+\ga_{N+1}^2(\tau)
\right.\right.\\{}&\left.\left.
+\be_N^2(\tau)+2\be_N(\tau)\be_{N-1}(\tau)
+\be_{N-1}^2(\tau)\right]-\frac{1}{2}\right\}d\tau,
\end{aligned}
\end{equation}
where $\ga_n(\tau)$, $\be_n(\tau)$, $n=N-1,\,N,\,N+1,$ are the
recurrence coefficients   
for orthogonal polynomials with respect to the weight
$e^{-NV(z;\tau)}$,  and
\begin{equation}\label{fin5}
F_N^{\Gauss}=-\frac{1}{N^2}\,\ln
\left(\frac{(2\pi)^{N/2}}{(2N)^{N^2/2}} 
\prod_{n=1}^N n!\right)
\end{equation}
is the free energy of the Gaussian ensemble.
\end{theo}

\begin{proof}
From Proposition \ref{finite_n} we obtain that
\begin{equation}\label{fin6}
\ga_N^2(\tau)\left[\ga_{N-1}^2(\tau)+\ga_{N+1}^2(\tau)
+\be_N^2(\tau)+2\be_N(\tau)\be_{N-1}(\tau)
+\be_{N-1}^2(\tau)\right]-\frac{1}{2}
=O(\tau^{-1/2})\,,
\end{equation}
hence the integral in (\ref{fin4}) converges. In additon,
$F_N(\infty)$ is the free energy for the Gaussian ensemble.
Since for the Gaussian ensemble,
\begin{equation}\label{fin7}
h_n=h_0\ga^2_1\ldots \ga^2_n=\frac{\sqrt{\pi}}{\sqrt N}\,\frac{1}{(2N)}\ldots 
\frac{n}{(2N)}=\frac{\sqrt{\pi}}{\sqrt N}\,\frac{n!}{(2N)^n},
\end{equation}
we obtain from (\ref{f1}) that
\begin{equation}\label{fin8}
Z_N(\infty)=N!\prod_{n=0}^{N-1}
\left[\frac{\sqrt{\pi}}{\sqrt N}\,\frac{n!}{(2N)^n}\right]
=\frac{(2\pi)^{N/2}}{(2N)^{N^2/2}}
\prod_{n=1}^N n!\,,
\end{equation}
so that $F_N(\infty)=F_N^{\Gauss}$. Denote the function on the right
in (\ref{fin4}) by $\wt F_N(t)$. Then by Proposition \ref{diff_t},
\begin{equation}\label{fin9}
\begin{aligned}
\frac{\partial^{2}F_N(t)}{\partial t^2}
&=-\frac{1}{t^2}\left\{\ga_N^2(t)\left[\ga_{N-1}^2(t)+\ga_{N+1}^2(t)
+\be_N^2(t)+2\be_N(t)\be_{N-1}(t)+\be_{N-1}^2(t)\right]-\frac{1}{2}\right\}\\
{}&=\frac{\partial^{2}\wt
  F_N(t)}{\partial t^2} ,
\end{aligned}
\end{equation}
hence $F_N(t)-\wt F_N(t)=at+b$. Since $F_N(\infty)=\wt
F_N(\infty)=F_N^{\Gauss}$, 
we obtain that $a=b=0$, hence $F_N(t)=\wt F_N(t)$. Theorem \ref{F_N} is proven.
\end{proof}

\section{Analyticity of the Equilibrium Measure for a Regular $V$}\label{1cut_regular}

This section is auxiliary.  We prove in this section a general
theorem on the analyticity of the equilibrium measure with respect to
perturbations of a regular $V$. The proof will follow directly
from a result of Kuijlaars and McLaughlin \cite{KM}, that the Jacobian
of the map of the end-points of the equilibrium measure to the
integrals $\{T_j,N_k\}$ is nonzero. Let us introduce the main
definitions. We will assume that $V(z)$ is a real analytic function
satisfying the growth condition,
\begin{equation}\label{an:1}
\lim_{|x|\to\infty}\frac{V(x)}{\log |x|}=\infty .
\end{equation}
The weighted energy of a Borel probability measure $\mu$ on the line      
is 
\begin{equation}\label{an:2}
I_V(\nu)=-\iint_{\R^2} \log|x-y| d\nu(x) d\nu(y)+\int_{\R^1}V(x)
d\nu(x). 
\end{equation}
There exists a unique equilibrium probability measure $\nu_{\eq}$,
which minimizes the functional $I_V(\nu)$,
\begin{equation}\label{an:3}
I_V(\nu_{\eq})=\min\,\{I_V(\nu):\; \nu\ge 0,\;\int_{\R^1} d\nu=1\}.
\end{equation}
As shown in \cite{DKM}, the equilibrium measure is absolutely continuous
and it is supported by a finite number of intervals,  
\begin{equation}\label{an:3a}
\textrm{supp}\, \nu_{\eq}=\cup_{i=1}^q [a_i,b_i].
\end{equation}
The density of $\nu_{\eq}$ on the support is given by the formula
\begin{equation}\label{an:4}
\rho(x)=\frac{1}{2\pi i}h(x)\sqrt{R_+(x)},\qquad x\in \cup_{i=1}^q [a_i,b_i],
\end{equation}
where the function $h$ is real analytic and
\begin{equation}\label{an:5}
R(z)=\prod_{i=1}^q[(z-a_i)(z-b_i)].
\end{equation}
For $\sqrt{R(z)}$ the principal sheet is taken, with cuts on 
$\cup_{i=1}^q [a_i,b_i]$, and $\sqrt{R_+(x)}$ means the value on the
upper cut. 
The equilibrium measure, $\nu=\nu_{\eq}$, satisfies the following
variational 
conditions: there exists a real constant $l=l_V$ such that
\begin{align}\label{an:6a}
L\nu(x)-\frac{V(x)}{2}&=l,\quad x\in \cup_{i=1}^q [a_i,b_i],\\
\label{an:6b}
L\nu(x)-\frac{V(x)}{2}&\le l,\quad x\in \R^1\setminus\cup_{i=1}^q [a_i,b_i],
\end{align}
where 
\begin{equation}\label{an:7}
L\nu(z)=\int_{\cup_{i=1}^q [a_i,b_i]}\log|z-x| d\nu(x),\quad z\in\C.
\end{equation}
Set
\begin{equation}\label{an:7a}
\om(z)=\int_{\R^1}\frac{d\nu(x)}{z-x},\quad z\in\C.
\end{equation}
Then 
\begin{equation}\label{f_e1}
\om(z)=\frac{1}{2}V'(z)-\frac{1}{2} h(z)\sqrt{R(z)},
\end{equation}
see, e.g., \cite {DKMVZ}. This implies that
\begin{equation}\label{f_e2}
h(z)=\frac{1}{2\pi i}\oint_{\Ga}\frac{V'(s)}{\sqrt{R(s)}\,(s-z)}\,ds,
\end{equation}
where $\Ga$ is a positively oriented contour in $\Omega$ around
$\{z\}\cup_{i=1}^q [a_i,b_i]$. Also, for $j=0,\dots,q$,
\begin{equation}\label{f_e3}
T_j\equiv \frac{1}{2\pi i}\oint_{\Ga}\frac{V'(z)z^j}{\sqrt{R(z)}}\,dz=2\de_{jq}.
\end{equation}
where $\Ga$ is a positively oriented contour in $\Omega$ around
$\cup_{i=1}^q [a_i,b_i]$.

A real analytic $V$ is called regular if
\begin{enumerate}
\item{ Inequality (\ref{an:6b}) is strict for all $x
\in \R^1\setminus\cup_{i=1}^q [a_i,b_i]$,}
\item{ $h(x)>0$ for all $x\in \cup_{i=1}^q [a_i,b_i]$.}
\end{enumerate}
Otherwise $V$ is called singular.
We formulate now the main result of this section.

\begin{theo}\label{analyticity} Suppose $V(z;t)$, $t\in[-t_0,t_0]$,
  $t_0>0$, 
is a one-parameter family of real analytic functions such that
\begin{enumerate}
\item[(a)]{there exists a domain $\Omega\subset\C$ such that
    $\R\subset\Omega$ and 
such that $V(z;t)$ is analytic on $\Omega\times[-t_0,t_0]$,}
\item[(b)]{$V(x,t)$ satisfies the uniform growth condition,
\begin{equation}\label{an:8} 
\lim_{|x|\to\infty}\frac{\min\{V(x;t):\;|t|\le t_0\}}{\log|x|}=\infty,
\end{equation}}
\item[(c)]{$V(z;0)$ is regular.}
\end{enumerate}
Then there exists $t_1>0$ such that if $t\in[-t_1,t_1]$, then
\begin{enumerate}
\item{ $V(z;t)$ is regular,}
\item{ the number $q$ of the intervals of the support of the
    equilibrium measure of $V(z;t)$ is independent of $t$, and}
\item{the end-points of the support intervals, $a_i(t),b_i(t)$,
    $i=1,\dots,q$, are real analytic 
    functions on $[-t_1,t_1]$.} 
\end{enumerate}   
\end{theo}

\begin{proof}
The regularity of $V(z;t)$ and $t$-independence of $q$ are proved in
\cite{KM}. To prove the analyticity consider the system of equations
on $\{a_i,b_i,\;i=1,\dots,q\}$,
\begin{equation}\label{an:9}
T_j=2\de_{kq},\quad j=0,1,\dots,q;
\qquad N_k=0,\quad k=1,\dots,q-1, 
\end{equation}
where $T_j$ is defined in (\ref{f_e3}) and
\begin{equation}\label{an:11}
N_k=\frac{1}{2\pi i}\oint_{\Ga_k} h(z)\sqrt{R(z)}\,dz,
\end{equation}
where $\Ga_k$ is a positively oriented contour around $[b_k,a_{k+1}]$,
which lies in a small neighborhood of $[b_k,a_{k+1}]$, so that
$\Ga_k\subset\Omega$ and it does not contain the other end-points.
In (\ref{an:11}) it is assumed that
the function $\sqrt{R(z)}$ is defined in such a way that it has a cut
on $[b_k,a_{k+1}]$.  
As shown in \cite{KM}, the Jacobian of the map 
$\{ [a_i,b_i]\}\to \{T_j,N_k\}$ is nonzero. The functions
$T_j,N_k$ are analytic with respect to $a_i,b_i$ and $t$.
By the implicit function
theorem, this implies the analyticity of $a_i(t),b_i(t)$. Theorem
\ref{analyticity} is proved.
\end{proof}

When applied to a polynomial $V$, Theorem \ref{analyticity} gives the
following result.

\begin{cor}
Suppose $V(z)=v_1z+\dots+v_{2d}z^{2d}$, $v_{2d}>0$, is $q$-cut
regular. Then for any $p\le 2d$ there exists $t_1>0$ such that for any
$t\in[-t_1,t_1]$, 
\begin{enumerate}
\item{ $V(z;t)=V(z)+tz^p$ is $q$-cut regular,}
\item{the end-points  of the support intervals, $a_i(t),b_i(t)$,
    $i=1,\dots,q$, are real analytic functions on $[-t_1,t_1]$.} 
\end{enumerate}   
\end{cor}

Theorem \ref{analyticity} can be applied to prove the analyticity of
the ($N=\infty$)-free energy,
\begin{equation}\label{an:12}
F=\lim_{N\to\infty}-\frac{1}{N^2}\ln Z_N.
\end{equation}
If $V$ is real analytic satisfying growth condition (\ref{an:1}), then
the limit on the right exists, see [Joh], and 
\begin{equation}\label{an:12a}
F=I_V(\nu_{\eq}).
\end{equation}

\begin{theo} \label {a_F}
Under the conditions of Theorem \ref{analyticity},
the free energy $F=F(t)$ is analytic on $[-t_1,t_1]$.
\end{theo}

\begin{proof}
The density of the equilibrium measure has form (\ref{an:4}), where
$h(x)$ is a real analytic function, which is found by formula (\ref{f_e2}).
 By Theorem \ref{analyticity} the end-points
of the support of $\nu_{\eq}$ depend analytically on $t$, hence
(\ref{f_e2}) implies that $h$ depends analytically on $t$, and,
therefore, $\nu_{eq}$ depends analytically on $t$. Formula
(\ref{an:12}) implies the analyticity of $F$. Theorem
\ref{a_F} is proved.
\end{proof}

Theorem \ref{a_F} implies that the critical points of the random
matrix model, the points
of nonanalyticity of the free energy, are at singular $V$ only.

\section{Asymptotic Expansion of the Recurrence Coefficients 
for a One-Cut Regular Polynomial $V$} \label{rec_coeff}

In this section we will assume that $V(z)$ is a polynomial,
which possesses a
one-cut regular equilibrium measure.
The equilibrium measure is one-cut means that its support
consists of one interval $[a,b]$, and if it is one-cut regular
then
\begin{equation}\label{rn:2}
d\nu_{\eq}(x)=\frac{1}{2\pi}h(x)\sqrt{(b-x)(x-a)},\quad x\in[a,b],
\end{equation}
where $h(x)$ is a polynomial such that $h(x)>0$ for all 
real $x$ (see the work of Deift, Kriecherbauer and
McLaughlin \cite{DKM}). For the sake of brevity, we will say
that $V(x)$ is one-cut regular if
its equilibrium measure is one-cut regular. 

As shown by Kuijlaars and McLaughlin \cite{KM}, if $V(x)$ is one-cut
regular then there exists $\ep>0$ such that for any $s$ in the
interval $1-\ep\le s\le 1+\ep$, the polynomial $s^{-1}V(x)$ is one-cut
regular, and the end-points, $a(s),b(s)$, are analytic functions
of $s$ such that $a(s)$ is decreasing and $b(s)$ is increasing. 
(In fact, the result of Kuijlaars and McLaughlin is much more
general and it includes multi-cut $V$ as well.) 

\begin{prop}  \label{rec1}
Suppose $V(x)$ is one-cut regular.
Then there exists $\eps>0$ such 
that for all $n$ in the interval
\begin{equation}\label{rec:1}
1-\eps\le \frac{n}{N}\le 1+\eps,
\end{equation}
 the recurrence coefficients  admit the uniform asymptotic
representation,   
\begin{equation}\label{rec:2}
\ga_n= \ga\left(\frac{n}{N}\right)+O(N^{-1}),
\quad
\be_n= \be\left(\frac{n}{N}\right)+O(N^{-1}).
\end{equation}
The functions $\ga(s),\,\be(s)$ are expressed as
\begin{equation}\label{rec:3}
\ga(s)=\frac{b(s)-a(s)}{4},\quad \be(s)=\frac{a(s)+b(s)}{2},
\end{equation}
where $[a(s),b(s)]$ is the
support of the equilibrium measure for the polynomial $s^{-1}V(x)$.
\end{prop}

\begin{proof} For $n=N$ the result follows from \cite{DKMVZ}. For a general
$n$, we can write $NV=ns^{-1}V$, $s=n/N$, and the result follows from 
the mentioned above result from \cite{KM}, that $s^{-1}V$ is one-cut
regular, and from   
\cite{DKMVZ}. The uniformity of the estimate of the error term follows
from the result from \cite{KM} on the analytic dependence of the equilibrium
measure of $s^{-1}V$ on $s$
and from the proof in \cite{DKMVZ}. 
\end{proof}

We can now formulate the main result of this section.

\begin{theo} \label{asympt} Suppose that $V(x)$ is a one-cut regular
polynomial.
Then there exists $\eps>0$ such that for all $n$ in the interval (\ref{rec:1}),
 the recurrence coefficients  
admit the following
uniform asymptotic expansion as $N\to\infty$ in powers of $N^{-2}$:
\begin{equation}\label{rn:4}
\ga_n\sim \ga\left(\frac{n}{N}\right)+\sum_{k=1}^\infty N^{-2k}
f_{2k}\left(\frac{n}{N}\right)\,,\quad 
\be_n\sim \be\left(\frac{n+\frac{1}{2}}{N}\right)+\sum_{k=1}^\infty N^{-2k}
g_{2k}\left(\frac{n+\frac{1}{2}}{N}\right)\,, 
\end{equation}
where  $f_{2k}(s)$, $g_{2k}(s)$, $k\ge 1$, are analytic functions 
on $[1-\eps,1+\eps]$.
\end{theo}

\begin{proof} 
Let us remind that
the proof in \cite{DKMVZ} of the asymptotic formula for  the recurrence
  coefficients is based on a reduction of the Riemann-Hilbert (RH) problem
for orthogonal polynomials to a RH problem in which all the
jumps are of the order of $N^{-1}$. By iterating the reduced RH
problem, one obtains an asymptotic expansion of the recurrence
coefficients,
\begin{equation}\label{th1}
\ga_N\sim \ga+\sum_{k=1}^\infty N^{-k}
f_k\,,\quad 
\be_N\sim \be+\sum_{k=1}^\infty N^{-k}
\hat g_k\,. 
\end{equation}
For a general $n$, let us write $NV=ns^{-1}V$, $s=n/N$. Then, as shown 
in \cite{KM}, the equilibrium meassure of $s^{-1}V$ is one-cut regular
and it depends analytically on $s$ in the interval $[1-\ep,1+\ep]$.
As follows from the iterations of the reduced RH problem, the
coefficients
$f_k$, $\hat g_k$ are expressed analytically in terms of the equilibrium
measure and hence they analytically depend on $n/N$, so that
\begin{equation}\label{rn:5}
\ga_n\sim \ga\left(\frac{n}{N}\right)+\sum_{k=1}^\infty N^{-k}
f_k\left(\frac{n}{N}\right)\,,\quad 
\be_n\sim \be\left(\frac{n}{N}\right)+\sum_{k=1}^\infty N^{-k}
\hat g_k\left(\frac{n}{N}\right)\,, 
\end{equation}
where $f_k(s),\hat g_k(s)$ are analytic functions on $[1-\ep,1+\ep]$.
We can rewrite the expansion of $\be_n$ in the form
\begin{equation}\label{rn:6}
\ga_n\sim \ga\left(\frac{n}{N}\right)+\sum_{k=1}^\infty N^{-k}
f_{k}\left(\frac{n}{N}\right)\,,\quad 
\be_n\sim \be\left(\frac{n+\frac{1}{2}}{N}\right)+\sum_{k=1}^\infty N^{-k}
g_{k}\left(\frac{n+\frac{1}{2}}{N}\right)\,, 
\end{equation}
where $g_k(s)$ are analytic on $[1-\ep,1+\ep]$.
What we have to prove is that $f_k=g_k=0$ for odd $k$. This will be
done by using the string equations.

Recall the string equations for the recurrence coefficients,
\begin{equation}\label{rn:3}
\ga_n [V'(Q)]_{n,n-1}=\frac{n}{N}\,,\quad [V'(Q)]_{nn}=0.
\end{equation}
where $[V'(Q)]_{nm}$ is the element $(n,m)$ of the matrix $V'(Q)$.
We have, in particular, that
\begin{equation}\label{smk}
\begin{aligned}
{}&[Q]_{n,n-1}=\ga_n,\quad [Q]_{nn}=\be_n;\\
{}&[Q^2]_{n,n-1}=\be_{n-1}\ga_n+\be_n\ga_n,\quad
[Q]_{nn}=\ga_n^2+\be_n^2+\ga_{n+1}^2;\\ 
{}&[Q^3]_{n,n-1}=\ga_{n-1}^2\ga_n+\ga_n^3+\ga_n\ga_{n+1}^2
+\be_{n-1}^2\ga_n+\be_{n-1}\be_n\ga_n+\be_n^2\ga_n,\\ 
{}& [Q^3]_{nn}=\be_{n-1}\ga_n^2+2\be_n\ga_n^2
+2\be_n\ga_{n+1}^2+\be_{n+1}\ga_{n+1}^2+\be_n^3,
\end{aligned}
\end{equation}
and so on.

\begin{lem}\label{symm_rec} For any $k\ge 1$, the expression of 
$[Q^k]_{n,n-1}$ in terms of $\ga_j,\be_j$
is invariant with respect to the change of variables 
\begin{equation}\label{tau0}
\sigma_0=\{\ga_j\to\ga_{2n-j},\; \be_j\to \be_{2n-j-1},\;j=0,1,
2,\ldots\},  
\end{equation}
provided $n>j+k$.
Similarly, the expression of 
$[Q^k]_{n,n}$ in terms of $\ga_j,\be_j$
is invariant with respect to the change of variables 
\begin{equation}\label{tau1}
\sigma_1=\{\ga_{n+j}\to\ga_{n-j+1},\; \be_{n+j}\to \be_{n-j},\;j=0,1,
2,\ldots\},  
\end{equation}
provided $n>j+k$.
\end{lem}

\begin{proof} Observe that the matrix $Q^k$ is symmetric.
By the rule of multiplication of matrices,
\begin{equation}\label{rn:7}
[Q^k]_{n,n-1}=\sum Q_{n,j_1}\dots Q_{j_{k-1},n-1}
=\sum Q_{n,\sigma(j_{k-1})}\dots Q_{\sigma(j_1),n-1},
\end{equation}
where $\sigma(j)\equiv 2n-j-1$. Observe that $\sigma(n)=n-1$,
$\sigma(\sigma(j))=j$ and
\[
Q_{jj}=\be_j,\quad Q_{\sigma(j),\sigma(j)}=\be_{2n-j-1};\qquad
Q_{j,j-1}=\ga_j,\quad Q_{\sigma(j),\sigma(j-1)}=\ga_{2n-j}
\]
This proves the invariance of $[Q^k]_{n,n-1}$ with respect
to $\sigma_0$. The invariance of $[Q^k]_{nn}$ with respect
to $\sigma_1$ is established similarly. Lemma \ref{symm_rec}
is proved. 
\end{proof} 

Since $V'(Q)$ is a linear combination of powers of $Q$, we
obtain the following corrolary of Lemma \ref{symm_rec}.

\begin{cor} The expression of $\ga_n[V'(Q)]_{n,n-1}$ (respectively,
 $[V'(Q)]_{nn}$) in terms
of $\ga_j,\be_j$ is invariant with respect to the change of variables
$\sigma_0$ (respectively, $\sigma_1$).
\end{cor}

Let us 
\begin{enumerate}
\item{substitute asymptotic 
expansions (\ref{rn:6}) into equations (\ref{rn:3})
and expand into powers series in $N^{-1}$,}
\item{expand $\gamma\left(\frac{n+j}{N}\right)$, 
$f_k\left(\frac{n+j}{N}\right)$,
$\beta\left(\frac{n+\frac{1}{2}+j}{N}\right)$, 
$g_k\left(\frac{n+\frac{1}{2}+j}{N}\right)$ 
in the Taylor series at $s=\frac{n}{N}$,}
\item{equate coefficients at powers of $N^{-1}$.}
\end{enumerate}
 This gives a system of equations on $\ga,\be,f_k,g_k$. 
 The zeroth order equations read 
\begin{equation}\label{th2}
\ga [V'(Q_0)]_{n,n-1}=s\,,\quad [V'(Q_0)]_{nn}=0,
\end{equation}
where $Q_0$ is a constant infinite Jacobi (tridiagonal)
matrix, such that
\begin{equation}\label{jac1}
[Q_0]_{nn}=\be,\quad [Q_0]_{n,n-1}=[Q_0]_{n-1,n}=\ga,\quad n\in\Z.
\end{equation}
Equations (\ref{th2}) are written as
\begin{equation}\label{jac2}
A(\ga,\be)=s,\quad B(\ga,\be)=0,
\end{equation}
where
\begin{equation}\label{jac3}
\begin{aligned}
A(\ga,\be)&=\ga\sum_{j=2}^{2d} jv_j\sum_{k=0}^{\left[\frac{j-2}{2}\right]}
\be^{j-2k-2}\ga^{2k+1}\binom{j-1}{2k+1}\binom{2k+1}{k}\,,\\
B(\ga,\be)&=\sum_{j=1}^{2d} jv_j\sum_{k=0}^{\left[\frac{j-1}{2}\right]}
\be^{j-2k-1}\ga^{2k}\binom{j-1}{2k}\binom{2k}{k}\,.
\end{aligned}
\end{equation}
Observe that $\ga,\be$ given in (\ref{rec:3}) solve equations 
(\ref{jac2}). The $k$-th order equations for $k\ge 1$ have the form
\begin{equation}\label{f_o0}
\begin{aligned}
{}& \frac{\partial A(\ga,\be)}{\partial\ga} f_k
+\frac{\partial A(\ga,\be)}{\partial\be}g_k=p,\\
{}&\frac{\partial B(\ga,\be)}{\partial\ga} f_k
+\frac{\partial B(\ga,\be)}{\partial\be} g_k=q,
\end{aligned}
\end{equation}
where $p,q$ are expressed in terms of the previous coefficients,
$\ga,\be,f_1,g_1,\dots,f_{k-1},g_{k-1}$, and their derivatives.
Here the  partial derivatives on the left are evaluated at $\ga,\be$
given in (\ref{rec:3}). 

\begin{lem} \label{first_order}
The first order equations are
\begin{equation}\label{f_o1}
\begin{aligned}
{}& \frac{\partial A(\ga,\be)}{\partial\ga} f_1
+\frac{\partial A(\ga,\be)}{\partial\be}g_1=0,\\
{}&\frac{\partial B(\ga,\be)}{\partial\ga} f_1
+\frac{\partial B(\ga,\be)}{\partial\be} g_1=0.
\end{aligned}
\end{equation}
\end{lem}

 \begin{proof} Observe that the terms with $f_1,g_1$ are the only
 first order  terms which appear at step (1) above. All the other terms appear at
   step (2), in the expansion of $\gamma\left(\frac{n+j}{N}\right)$, 
$f_k\left(\frac{n+j}{N}\right)$,
$\beta\left(\frac{n+\frac{1}{2}+j}{N}\right)$, 
$g_k\left(\frac{n+\frac{1}{2}+j}{N}\right)$ 
in the Taylor series at $s=\frac{n}{N}$. Consider any monomial 
on the left   in the first equation in (\ref{rn:3}), 
\[
C\ga_{n+j_1}\dots\ga_{n+j_p}\be_{n+l_1}\dots\be_{n+l_q}.
\]
By lemma \ref{symm_rec}, there is a partner to this term
of the form
\[
C\ga_{n-j_1}\dots\ga_{n-j_p}\be_{n-l_1-1}\dots\be_{n-l_q-1}.
\]
When we substitute expansions (\ref{rn:6}), we obtain
\[
C\left(\ga(s+\frac{j_1}{N})+\dots\right)\dots
\left(\ga(s+\frac{j_p}{N})+\dots\right)
\left(\be(s+\frac{l_1+\frac{1}{2}}{N})+\dots\right)
\dots \left(\be(s+\frac{l_q+\frac{1}{2}}{N})+\dots\right)
\]
and
\[
C\left(\ga(s-\frac{j_1}{N})+\dots\right)\dots
\left(\ga(s-\frac{j_p}{N})+\dots\right)
\left(\be(s-\frac{l_1+\frac{1}{2}}{N})+\dots\right)
\dots \left(\be(s-\frac{l_q+\frac{1}{2}}{N})+\dots\right)
\]
for the partner. When we expand these expressions in powers
of $N^{-1}$, the first order terms cancel
each other in the sum of
the partners (in fact, all the odd
terms cancel). This proves the first equation in (\ref{f_o1}).
The second one is proved similarly. 
Lemma \ref{first_order} is proved.  
\end{proof}

Lemma \ref{first_order} implies that $f_1(s)=g_1(s)=0$ for all $s$
such that
\begin{equation}\label{th11}
\det \begin{pmatrix}
\frac{\partial A(\ga,\be)}{\partial\ga} &
\frac{\partial A(\ga,\be)}{\partial\be} \\
\frac{\partial B(\ga,\be)}{\partial\ga} &
\frac{\partial B(\ga,\be)}{\partial\be}
\end{pmatrix} 
\not=0,
\end{equation}
where all the partial derivatives are evaluated at $\ga(s),\be(s)$ given in
(\ref{rec:3}).

\begin{lem}\label{any_order} 
If for a given $s\in[1-\ep,1+\ep]$, 
condition (\ref{th11}) holds, then all odd coefficients
$f_{2k+1}(s),g_{2k+1}(s)$ are zero.
\end{lem}

\begin{proof} 
By Lemma \ref{first_order} $f_1(s)=g_1(s)=0$. If we
  consider terms of the order of $N^{-3}$ then we obtain the equations
\begin{equation}\label{f_o2}
\begin{aligned}
{}& \frac{\partial A(\ga,\be)}{\partial\ga} f_3
+\frac{\partial A(\ga,\be)}{\partial\be}g_3=0,\\
{}&\frac{\partial B(\ga,\be)}{\partial\ga} f_3
+\frac{\partial B(\ga,\be)}{\partial\be} g_3=0.
\end{aligned}
\end{equation}
Indeed,
the same argument as in Lemma \ref{first_order} proves that all
other terms of the third order cancel out. Since condition (\ref{th11})
holds, it implies that $f_3(s)=g_3(s)=0$. By continuing this argument
we prove that all odd $f_{2k+1}(s),g_{2k+1}(s)$ vanish. Lemma \ref{any_order} is
proved. 
\end{proof}

\begin{lem}\label{Jacobian}
Condition (\ref{th11}) holds for all $s\in[1-\ep,1+\ep]$.
\end{lem}

\begin{proof}
 By differentiating equations
(\ref{jac2}) in $s$ we obtain that
\begin{equation}\label{f_o3}
\begin{aligned}
{}& \frac{\partial A(\ga,\be)}{\partial\ga} \frac{\partial
  \ga}{\partial s} 
+\frac{\partial A(\ga,\be)}{\partial\be}\frac{\partial
  \be}{\partial s}=1,\\
{}&\frac{\partial B(\ga,\be)}{\partial\ga} \frac{\partial
  \ga}{\partial s}
+\frac{\partial B(\ga,\be)}{\partial\be} \frac{\partial
  \be}{\partial s}=0.
\end{aligned}
\end{equation} 
 By differentiating equations
(\ref{jac2}) in $t_1$ we obtain that
\begin{equation}\label{f_o4}
\begin{aligned}
{}& \frac{\partial A(\ga,\be)}{\partial\ga} \frac{\partial
  \ga}{\partial t_1}
+\frac{\partial A(\ga,\be)}{\partial\be}\frac{\partial
  \be}{\partial t_1}=0,\\
{}&\frac{\partial B(\ga,\be)}{\partial\ga} \frac{\partial
  \ga}{\partial t_1}
+\frac{\partial B(\ga,\be)}{\partial\be} \frac{\partial
  \be}{\partial t_1}=-1.
\end{aligned}
\end{equation}
By rewriting equations (\ref{f_o3}), (\ref{f_o4}) in the matrix form,
we obtain that 
\begin{equation}\label{f_o5}
\begin{pmatrix}
\frac{\partial A(\ga,\be)}{\partial\ga} &
\frac{\partial A(\ga,\be)}{\partial\be} \\
\frac{\partial B(\ga,\be)}{\partial\ga} &
\frac{\partial B(\ga,\be)}{\partial\be}
\end{pmatrix} 
\begin{pmatrix}
\frac{\partial\ga}{\partial s} 
& \frac{\partial  \ga}{\partial t_1} \\
\frac{\partial\ga}{\partial s} 
& \frac{\partial  \ga}{\partial t_1}
\end{pmatrix}
=\begin{pmatrix}
1 & 0 \\
0 & -1
\end{pmatrix}.
\end{equation}
Since 
\[
\det\begin{pmatrix}
1 & 0 \\
0 & -1
\end{pmatrix}=-1\not=0,
\]
this implies (\ref{th11}). Lemma \ref{Jacobian} is proved.
\end{proof} 

From Lemmas \ref{any_order} and \ref{Jacobian} we obtain that the odd
coefficients $f_{2k+1},g_{2k+1}$ vanish. Theorem \ref{asympt} is
proved.
\end{proof}

\section{Asymptotic Expansion of the Free Energy 
for a One-Cut Regular $V$}\label{free_energy}

We have the following extension of Theorem \ref{asympt}.

\begin{theo} \label{asympt_t} Suppose that $V(z;t)$,
  $t\in[-t_0,t_0]$, $t_0>0$, is  a one-parameter analytic family of polynomials
of degree $2d$, such that $V(z;0)$ is one-cut regular.
Then there exist $t_1>0$ and $\ep>0$ such that for all $t\in[-t_1,t_1]$
and all $n\in[(1-\ep)N,(1+\ep)N]$,
 the recurrence coefficients corresponding to $V(z;t)$, 
admit the following
uniform asymptotic expansion as $N\to\infty$:
\begin{equation}\label{rn:4a}
\ga_n\sim \ga\left(\frac{n}{N};t\right)+\sum_{k=1}^\infty N^{-2k}
f_{2k}\left(\frac{n}{N};t\right)\,,\quad 
\be_n\sim \be\left(\frac{n+\frac{1}{2}}{N};t\right)+\sum_{k=1}^\infty N^{-2k}
g_{2k}\left(\frac{n+\frac{1}{2}}{N};t\right)\,, 
\end{equation}
where  $\ga(s;t)$, $\be(s;t)$,
$f_{2k}(s;t)$, $g_{2k}(s;t)$, $k\ge 1$, are analytic functions of $s,t$
on $[1-\eps,1+\eps]\times[-t_1,t_1]$.
\end{theo}

\begin{proof} Theorem \ref{analyticity} implies that 
the equilibrium measure of
$V(z;t)$ is analytic in $t\in[-t_1,t_1]$ (see the proof
of Theorem \ref{a_F}). This implies the
analyticity of $\ga,\be$ in $s$ and $t$. From equations \ref{f_o0} we
obtain the analyticity of $f_k,g_k$, $k\ge 1$. By Theorem 
\ref{asympt} all odd $f_{2k+1},g_{2k+1}$ vanish. This proves Theorem \ref{asympt_t}.
\end{proof}

Let us return to the polynomial $V(z;t)=\tau_t V(z)$. 
We will assume the following hypothesis.

{\it Hypothesis R.} For all $t\ge 1$ the polynomial $\tau_t V(z)$ 
is one-cut regular.

\begin{theo}\label{expansion_Z}
If a polynomial $V(z)$ satisfies Hypothesis R, then its free energy
admits the asymptotic expansion,
\begin{equation}\label{e_Z1}
F_N-F_N^{\Gauss}\sim F+N^{-2}F^{(2)}+N^{-4}F^{(4)}+\ldots,
\end{equation}
where $F_N^{\Gauss}$ is defined in (\ref{fin5}). The leading term
of the asymptotic expansion is:
\begin{equation}\label{e_Z2}
F=\int_1^\infty \frac{1-\tau}{\tau^2}
\left[2\ga^4(\tau)+4\ga^2(\tau)\be^2(\tau)-\frac{1}{2}\right]
d\tau,
\end{equation}
where 
\begin{equation}\label{e_Z3}
\ga(\tau)=\frac{b(\tau)-a(\tau)}{4},\qquad
\be(\tau)=\frac{a(\tau)+b(\tau)}{2},
\end{equation}
and $[a(\tau),b(\tau)]$ is the support of the equilibrium measure
for the polynomial $V(z;\tau)$. The quantities $\ga=\ga(\tau)$, 
$\be=\be(\tau)$ solve the equations,
\begin{equation}\label{e_Z4}
A(\ga,\be;\tau)=1,\qquad B(\ga,\be;\tau)=0,
\end{equation}
where
\begin{equation}\label{ab1}
\begin{aligned}
A(\ga,\be;\tau)&=2\left(1-\frac{1}{\tau}\right)\ga^2+\ga\sum_{j=1}^{2d}
\frac{jv_j}{\tau^{j/2}}\sum_{k=0}^{\left[\frac{j-2}{2}\right]} 
\be^{j-2k-2}\ga^{2k+1}\binom{j-1}{2k+1}\binom{2k+1}{k}\,,\\
B(\ga,\be;\tau)&=2\left(1-\frac{1}{\tau}\right)\be+\sum_{j=1}^{2d}
\frac{jv_j}{\tau^{j/2}}\sum_{k=0}^{\left[\frac{j-1}{2}\right]} 
\be^{j-2k-1}\ga^{2k}\binom{j-1}{2k}\binom{2k}{k}\,.
\end{aligned}
\end{equation}
\end{theo}

\begin{proof} 
By applying Theorem \ref{asympt_t} to $\tau_t V(z)$, we obtain the
uniform asymptotic expansions,
\begin{equation}\label{e_Z5}
\ga_n(t)\sim \ga\left(\frac{n}{N};t\right)+\sum_{k=1}^\infty N^{-2k}
f_k\left(\frac{n}{N};t\right)\,,\quad 
\be_n(t)\sim
\be\left(\frac{n+\frac{1}{2}}{N};t\right)+\sum_{k=1}^\infty N^{-2k} 
g_k\left(\frac{n+\frac{1}{2}}{N};t\right)\,. 
\end{equation}
From (\ref{ab1}) with $\tau=t$, as $t\to\infty$, 
\begin{equation}\label{ab2}
A(\ga,\be;t)=2\ga^2+O(t^{-1/2}),\qquad 
B(\ga,\be;t)=2\be+O(t^{-1/2}),
\end{equation}
hence the solutions to system (\ref{e_Z4}) are
\begin{equation}\label{abt3}
\ga(t)=\frac{\sqrt 2}{2}+O(t^{-1/2}),\qquad \be(t)=O(t^{-1/2}).
\end{equation}
By differentiating equations (\ref{e_Z4}) in $\tau=t$ we obtain the 
equations,
\begin{equation}\label{f_o4a}
\begin{aligned}
{}& \frac{\partial A(\ga,\be;t)}{\partial\ga} \ga'(t)
+\frac{\partial A(\ga,\be;t)}{\partial\be}\be'(t)=p,\\
{}&\frac{\partial B(\ga,\be;t)}{\partial\ga} \ga'(t)
+\frac{\partial B(\ga,\be;t)}{\partial\be} \be'(t)=q,
\end{aligned}
\end{equation}
where $p,q$ are expressed in terms of $\ga(t),\be(t)$ and
$p,q=O(t^{-1/2})$. From this system of equations we obtain 
that $\ga'(t),\be'(t)=O(t^{-1/2})$. By differentiating 
equations (\ref{e_Z4}) many times we obtain the estimates
for $j\ge 1$,
\begin{equation}\label{abt4}
\ga^{(j)}(t)=O(t^{-1/2}),\qquad \be^{(j)}(t)=O(t^{-1/2}).
\end{equation}
From equations (\ref{f_o0}) we obtain the estimates on $f_k,g_k$,
\begin{equation}\label{abt5}
f_k^{(j)}(t)=O(t^{-1/2}),\qquad g_k^{(j)}(t)=O(t^{-1/2}),\qquad j\ge 0,
\end{equation}
and from the reduced RH problem, that for any $K\ge 0$,
\begin{equation}\label{e_Z5a}
\begin{aligned}
{}&\left|\ga_n(t)- \ga\left(\frac{n}{N};t\right)-\sum_{k=1}^K N^{-2k}
f_k\left(\frac{n}{N};t\right)\right|\le C(K)N^{-2K-2}t^{-1/2}\,,\\ 
{}&\left|\be_n(t)-
\be\left(\frac{n+\frac{1}{2}}{N};t\right)-\sum_{k=1}^K N^{-2k} 
g_k\left(\frac{n+\frac{1}{2}}{N};t\right)\right|\le
C(K)N^{-2K-2}t^{-1/2}\,.
\end{aligned}  
\end{equation}
(cf. the derivation of the estimates (\ref{gauss8}) and (\ref{gauss9}) in 
Appendix \ref{appendix}.)
Let us substitute expansions (\ref{e_Z5}) into (\ref{fin4}) and
expand the terms on the right in the Taylor series at $n/N=1$.
In this way we obtain the asymptotic expansion,
\begin{align}\label{FN:1a}
\Theta_{N}(\tau) &\equiv
\ga_N^2(\tau)\left[\ga_{N-1}^2(\tau)+\ga_{N+1}^2(\tau)
+\be_N^2(\tau)+2\be_N(\tau)\be_{N-1}(\tau)
+\be_{N-1}^2(\tau)\right]-\frac{1}{2}\\
\label{FN:1b}
{}&\sim
\Theta(\tau) + 
\sum_{k=1}^\infty N^{-k}\Theta^{(k)}(\tau)
\end{align}
where
\begin{equation}\label{FN:2}
\Theta(\tau)=2\ga^4(\tau)+4\ga^2(\tau)\be^2(\tau)-\frac{1}{2}
\end{equation}
and
\begin{equation}\label{FN:3}
\Theta^{(k)}(\tau)\le C(k)\tau^{-1/2},\qquad k\ge 1.
\end{equation}
Observe that the expression (\ref{FN:1a}) is invariant with respect to
the transformation
\begin{equation}\label{FN:4}
\sigma_0=\{\gamma_j\to \gamma_{2N-j},\;
\beta_j\to \beta_{2N-j-1}\}.
\end{equation}
Therefore, as in the proof of Lemmas \ref{first_order},
\ref{any_order}, we obtain that all odd $\Theta^{(2k+1)}=0$.
Theorem \ref{expansion_Z} is proved.
\end{proof}

The following parametric extension of Theorem \ref{expansion_Z} is
useful for applications. 

\begin{theo}\label{expansion_Z1}
Suppose $\{V(z;u),\;u\in[-u_0,u_0]\}$ is a one-parameter analytic
family of one-cut regular polynomials of degree $2d$ such that
the  polynomial $V(z;0)$ satisfies Hypothesis R. Then there exists
$u_1>0$ such that the coefficients
$F(u),F^{(2)}(u),F^{(4)}(u),\dots$ of the asymptotic expansion of the
free 
energy for $V(z;u)$ are analytic on $[-u_1,u_1]$.
\end{theo}

\begin{proof} The functions  are expressed in terms of
 integrals of finite combinations of the  functions
  $\ga^{(j)}(1;u),\be^{(j)}(1;u),
f^{(j)}_{2k}(1;u),g^{(j)}_{2k}(1;u)$. By Theorem \ref{asympt_t} these
  functions are analytic in $u$. By the same argument as in the proof
  of Theorem \ref{expansion_Z}, we obtain that they behave like
  $O(t^{-1/2})$ as $t\to\infty$. Therefore, the integrals expressing
  $F^{(2n})(u)$ converge and define an analytic function in
  $u$. Theorem \ref{expansion_Z1} is proved.
\end{proof} 

The following proposition is auxiliary: 
it gives first several terms of the asymptotic
expansion of the free energy for the Gaussian ensemble.

\begin{prop} The constant $F_N^{\Gauss}$ has the following expansion:
\begin{equation}\label{e_Z6}
\begin{aligned}
F_N^{\Gauss}&\sim \frac{\ln 2}{2} -\frac{3}{4}-\frac{\ln N}{N}
+(1-\ln(2\pi))\frac{1}{N}-\frac{5 \ln
  N}{12 N^2}-\left(\zeta'(-1)+\frac{\ln(2\pi)}{2}\right)\frac{1}{N^2}\\
{}&-\frac{1}{12N^3}+\frac{1}{240 N^4}+\frac{1}{360
  N^5}+O\left(\frac{1}{N^6}\right).
\end{aligned}
\end{equation}
\end{prop}

\begin{proof} This is obtained from (\ref{fin5}) with the help of
  MAPLE.
\end{proof}

\section{Exact Formula for the Free Energy for an Even $V$}
\label{exact_formula}

For an even $V$, $\be(\tau)=0$, and formula (\ref{e_Z2}) simplifies,
\begin{equation}\label{ev:1}
F=\int_1^\infty \frac{1-\tau}{\tau^2}
\left[2R^2(\tau)-\frac{1}{2}\right]
d\tau,
\end{equation}
where
\begin{equation}\label{ev:2}
R(\tau)=\ga^2(\tau).
\end{equation}
From (\ref{e_Z4}), (\ref{ab1}) we obtain that $R=R(\tau)$ solves the
equation 
\begin{equation}\label{ev:3}
2\left(1-\frac{1}{\tau}\right)R+\sum_{j=1}^{d}
jv_{2j} 
\binom{2j}{j}\left(\frac{R}{\tau}\right)^j=1. 
\end{equation}
Set
\begin{equation}\label{ef:1}
\xi(\tau)=\frac{R(\tau)}{\tau}\,.
\end{equation}
Then equation (\ref{ev:3}) is rewritten as
\begin{equation}\label{ef:3}
\tau =\frac{1}{2\xi}+1-\frac{1}{2}\,\sum_{j=1}^d \,v_{2j}\,
j\,\binom{2j}{j}\,\xi^{j-1}\equiv \tau(\xi)\,.
\end{equation}
From (\ref{ev:1}),
\begin{equation}\label{ef:4}
F=\int_1^\infty (1-\tau)\left[
2\xi^2(\tau)-\frac{1}{2\tau^2}\right] d\tau=\lim_{T\to\infty}\left[
\int_1^T 2(1-\tau)
\,\xi(\tau)^2 d\tau
+\frac{1}{2}\,\ln T-\frac{1}{2}\right]\,.
\end{equation}
The change of variable $\tau=\tau(\xi)$ reduces the
latter formula to
\begin{equation}\label{ef:5}
F=\lim_{T\to\infty}\left[
\int_{\xi(1)}^{\xi(T)} 2(1-\tau(\xi))
\,\xi^2 \tau'(\xi)\,d\xi
+\frac{1}{2}\,\ln T-\frac{1}{2}\right]\,,
\end{equation}
or, by (\ref{ef:3}), to
\begin{equation}\label{ef:6}
F=\lim_{T\to\infty}\frac{1}{2}\left[
\int_{\xi(1)}^{\xi(T)} \left(\frac{1}{\xi}
-\sum_{j=1}^d \,v_{2j}\,
j\,\binom{2j}{j}\,\xi^{j-1}\right)
\left(1+\sum_{j=1}^d \,v_{2j}\,
j(j-1)\,\binom{2j}{j}\,\xi^j\right)\,d\xi
+\ln T\right]\,.
\end{equation}
By (\ref{ef:1}), $\xi(1)=R(1)$. By (\ref{fin1}), 
\[
R(T)=\frac{1}{2}+O(T^{-1/2})\,,
\]
(since $n=N$), hence
\begin{equation}\label{ef:7}
\xi(T)=\frac{R(T)}{T}=\frac{1}{2T}+O(T^{-3/2})\,.
\end{equation}
When we distribute
on the right in (\ref{ef:6}), we have the following terms:
\begin{align}\label{ef:7a}
&(1)\quad\lim_{T\to\infty}\left[\int_{\xi(1)}^{\xi(T)}
\frac{1}{\xi}\,d\xi
+\ln T\right]=-\ln 2-\ln R(1)\,, \nonumber\\ 
&(2)\quad -\int_{\xi(1)}^0 
\sum_{j=1}^d \,v_{2j}\,
j\,\binom{2j}{j}\,\xi^{j-1} d\xi
=\sum_{j=1}^d v_{2j}\,
\binom{2j}{j}\,R(1)^j,
\nonumber\\
&(3)\quad\int_{\xi(1)}^0
\sum_{j=1}^d \,v_{2j}\,
j(j-1)\,\binom{2j}{j}\,\xi^{j-1}
=-\sum_{j=1}^d v_{2j}\,
(j-1)\binom{2j}{j}\,R(1)^j, 
\nonumber\\
&(4)\quad -\int_{\xi(1)}^0 
\left(\sum_{j=1}^d \,v_{2j}\,
j\,\binom{2j}{j}\,\xi^{j-1}\right)
\left(\sum_{k=1}^d \,v_{2k}\,
k(k-1)\,\binom{2k}{k}\,\xi^k\right)
d\xi
\nonumber\\
&\hskip 4cm
=\sum_{n=2}^{2d}
\left[\sum_{\substack{1\le j,k\le m\\j+k=n}}
v_{2j}v_{2k}\,jk(k-1)\binom{2j}{j}\binom{2k}{k}\right]
\frac{1}{n}\,R(1)^n\,,
\end{align}
hence (\ref{ef:6}) can be transformed into the following expression:
\begin{equation}\label{ef:8}
F=-\frac{\ln 2}{2}-\frac{\ln R}{2}\,
+\sum_{j=1}^d u_j\,R^j
+\sum_{n=2}^{2d} w_n\,R^n\,,
\end{equation}
where $R=R(1)$,
\begin{equation}\label{ef:9}
u_j=-\frac{j-2}{2}
\binom{2j}{j}v_{2j}
\end{equation}
and
\begin{equation}\label{ef:10}
w_n=\frac{1}{2n}\sum_{\substack{1\le j,k\le m\\j+k=n}}v_{2j}v_{2k}jk(k-1)
\binom{2j}{j}\binom{2k}{k}\,.
\end{equation}

\begin{example} The quartic polynomial,
\begin{equation}\label{ex0}
V(z)=\frac{z^4}{4}+tz^2, \qquad t>-1.
\end{equation}
The condition $t>-1$ is necessary and sufficient for one cut. By
(\ref{ev:3}), $R=R(1)>0$ solves the equation,
\begin{equation}\label{ef:11}
3R^2+2tR=1,
\end{equation}
hence
\begin{equation}\label{ef:12}
R=\frac{-t+\sqrt{t^2+3}}{3}\,.
\end{equation}
By (\ref{ef:9}), (\ref{ef:10}),
\begin{equation}\label{ef:13}
u_1=t,\quad u_2=0,\quad w_2=0,\quad
w_3=t,\quad w_4=\frac{9}{8}\,.
\end{equation}
Thus, by (\ref{ef:8}),
\begin{equation}\label{ef:14}
F=-\frac{\ln 2}{2}-\frac{\ln R}{2}\,
+tR+tR^3+\frac{9}{8}R^4.
\end{equation}
\end{example}

\section{One-Sided Analyticity for a
  Singular $V$}\label{one-sided}

In this section we will prove general results on the one-sided
analyticity for a singular $V$. We will assume the following
hypothesis.

{\it Hypothesis S.} $V(z;t)$, $t\in[0,t_0]$, is a one-parameter family
of real analytic functions  such that
\begin{enumerate}
\item[(a)]{there exists a domain $\Omega\subset\C$ such that
    $\R\subset\Omega$ and 
such that $V(z;t)$ is analytic on $\Omega\times[0,t_0]$,}
\item[(b)]{$V(x,t)$ satisfies the uniform growth condition,
\begin{equation}\label{os:1} 
\lim_{|x|\to\infty}\frac{\min\{V(x;t):\;0\le t\le t_0\}}{\log|x|}=\infty,
\end{equation}}
\item[(c)]{$V(z;t)$ is one-cut regular for $0< t\le t_0$,}
\item[(d)]{$V(z;0)$ is one-cut singular and $h(a)\not=0$,
$h(b)\not=0$, where $[a,b]$ is the support of the equilibrium measure
for $V(z;0)$.}
\end{enumerate}

\begin{theo}\label{os_analyticity}
Suppose $V(z;t)$ satisfies Hypothesis S. Then the end-points
$a(t),b(t)$
of the equilibrium measure for $V(z;t)$ are analytic on $[0,t_0]$.
\end{theo}

\begin{proof} Set
\begin{equation}\label{os:2}
T_j(a,b;t)=\frac{1}{2\pi i}\oint_{\Gamma}
  \frac{V'(z;t)z^j}{\sqrt{(z-a)(z-b)}}\, dz,
\end{equation}
where $\Gamma$ is a positively oriented closed contour around $[a,b]$ 
inside $\Omega$.  For $t\in[0,t_0]$, we have the following
equations on $a=a(t)$, $b=b(t)$:
\begin{equation}\label{os:3}
T_0(a,b;t)=0,\qquad T_1(a,b;t)=2.
\end{equation}
By differentiating (\ref{os:2}) we obtain that at $a=a(t)$, $b=b(t)$,
\begin{equation}\label{os:4}
\frac{\partial T_0(a,b;t)}{\partial a}
=\frac{1}{4\pi i}\oint_{\Gamma}\frac {V'(z)}{(z-a)\sqrt{R(z)}}dz
=\frac{1}{4\pi i}\oint_{\Gamma}\left(\frac{2\om(z)}{\sqrt{R(z)}}+h(z)\right)
\frac {1}{z-a}dz=\frac{h(a)}{2}.
\end{equation}
Similarly,
\begin{equation}\label{os:5}
\frac{\partial T_0(a,b;t)}{\partial b}=\frac{h(b)}{2},\qquad
\frac{\partial T_1(a,b;t)}{\partial a}=\frac{ah(a)}{2},\qquad
\frac{\partial T_1(a,b;t)}{\partial b}=\frac{bh(b)}{2}.
\end{equation}
Thus, the Jacobian,
\begin{equation}\label{os:6}
J=\det
\begin{pmatrix}
\frac{\partial T_0}{\partial a} & \frac{\partial T_0}{\partial b} \\ 
\frac{\partial T_1}{\partial a} & \frac{\partial T_1}{\partial b}
\end{pmatrix}=\frac{h(a)h(b)(b-a)}{4}\not=0.
\end{equation}
The function $T_0(a,b;t)$ is analytic in $a,b,t$, hence by the
implicit function theorem, $a(t),b(t)$ are analytic
on $[0,t_0]$. Theorem \ref{os_analyticity} is proved.
\end{proof}

\begin{cor}\label{cor_os}
Suppose $V(z;t)$ satisfies Hypothesis S. Then 
\begin{enumerate}
\item{the function $h(x;t)$ is analytic on $R^1\times [0,t_0]$,}
\item{the free energy $F(t)$ is analytic on $[0,t_0]$,}
\item{the functions $\ga(t),\be(t)$ are analytic on $[0,t_0]$.}
\end{enumerate}
\end{cor}

\begin{proof} The analyticity of $h$ follows from formula (\ref{f_e2}) 
and the one of $F$, from (\ref{an:12a}). Finally, the analyticity of 
$\ga(t),\be(t)$ follows from (\ref{rec:3}).
\end{proof}

Theorem \ref{os_analyticity} and Corollary \ref{cor_os} can be
extended to multi-cut $V$. We 
will say that $V$ is $q$-cut if the support of its equilibrium measure
consists of $q$ intervals, $[a_i,b_i]$, $i=1,\dots,q$.
 We will assume the following
hypothesis.

{\it Hypothesis \Sq.} $V(z;t)$, $t\in[0,t_0]$, is a one-parameter family
of real analytic functions  such that
\begin{enumerate}
\item[(a)]{there exists a domain $\Omega\subset\C$ such that
    $\R\subset\Omega$ and 
such that $V(z;t)$ is analytic on $\Omega\times[0,t_0]$,}
\item[(b)]{$V(x,t)$ satisfies the uniform growth condition,
\begin{equation}\label{os:1q} 
\lim_{|x|\to\infty}\frac{\min\{V(x;t):\;0\le t\le t_0\}}{\log|x|}=\infty,
\end{equation}}
\item[(c)]{$V(z;t)$ is $q$-cut regular for $0< t\le t_0$,}
\item[(d)]{$V(z;0)$ is $q$-cut singular (with the same $q$ as in (c))
    and $h(a_i)\not=0$, $h(b_i)\not=0$, $i=1,\dots,q$.}
\end{enumerate}
 
\begin{theo}\label{os_analyticity_q}
Suppose $V(z;t)$ satisfies Hypothesis \Sq. Then the end-points
$a_i(t),b_i(t)$
of the equilibrium measure for $V(z;t)$ are analytic on $[0,t_0]$.
\end{theo}

\begin{proof} Consider system of equations (\ref{an:9}) for
  $V=V(z;t)$. As shown in 
\cite{KM}, the Jacobian of the map
\begin{equation}\label{os:7} 
f:\;\{a_i,b_i,\;i=1,\dots,q\}
\to \{T_j,N_k,\;j=0,\dots,q;\;k=1,\dots,q-1\}
\end{equation}
at $\{a_i(t),b_i(t)\}$  is equal to
\begin{equation}\label{os:8} 
\begin{aligned}
\det&\left(\frac{\partial\{T_j,N_k\}}{\partial\{a_i,b_i\}}\right)
=\left(\prod_{i=1}^q \frac{\partial T_0}{\partial a_i} \frac{\partial
  T_0}{\partial b_i} \right)\pi^{-q+1}
\int_{b_1}^{a_2}\sqrt{R_+(x_1)}dx_1\dots
\int_{b_{q-1}}^{a_q}\sqrt{R_+(x_{q-1})}dx_{q-1}\\
{}&\times\det
\begin{pmatrix}
1 & 1 & \dots & 1 \\
a_1 & b_1 & \dots & b_q \\
\vdots & \vdots & \vdots & \vdots \\
a_1^q & b_1^q & \dots & b_q \\
(x_1-a_1)^{-1} & (x_1-b_1)^{-1} & \dots & (x_1-b_q)^{-1} \\
\vdots & \vdots & \vdots & \vdots \\
(x_{q-1}-a_1)^{-1} & (x_{q-1}-b_1)^{-1} & \dots & (x_{q-1}-b_q)^{-1}
\end{pmatrix}
\end{aligned}
\end{equation}
The determinant on the right is a mixture of a Vandermonde determinant
and a Cauchy determinant. As shown in \cite{KM}, it is equal to
\begin{equation}\label{os:9}
\frac{\prod_{j=1}^q\prod_{k=1}^q(b_k-a_j)
\prod_{1\le j<k\le q}(a_k-a_j)(b_k-b_j)
\prod_{1\le j<k\le q-1}(x_k-x_j)}
{(-1)^{q-1}\prod_{j=1}^{q-1}\prod_{k=1}^q (x_j-a_k)(x_j-b_k)},
\end{equation}
which is nonzero for
\begin{equation}\label{os:10}
a_1<b_1<x_1<a_2<\dots<b_{q-1}<x_{q-1}<a_q<b_q,
\end{equation}
and therefore has a fixed sign. Hence the multiple integral in
(\ref{os:8}) is nonzero. Now,
\begin{equation}\label{os:10a}
\frac{\partial T_0}{\partial a_i}
=\frac{1}{4\pi i}\oint_{\Gamma}\frac {V'(z)}{(z-a_i)\sqrt{R(z)}}dz
=\frac{1}{4\pi i}\oint_{\Gamma}\left(\frac{2\om(z)}{\sqrt{R(z)}}+h(z)\right)
\frac {1}{z-a_i}dz=\frac{h(a_i)}{2},
\end{equation}
and a similar formula holds for $\frac{\partial T_0}{\partial b_i}$.
Thus, the Jacobian (\ref{os:8}) is nonzero. The functions 
$\{T_j,N_k\}$ are analytic in $\{a_i,b_i\},t$, hence, by the
implicit function theorem, $\{a_i(t),b_i(t)\}$ are analytic
on $[0,t_0]$. Theorem \ref{os_analyticity_q} is proved.
\end{proof}

As a corollary of Theorem \ref{os_analyticity_q}, we obtain the
following results. 

\begin{cor}\label{cor_os_q}
Suppose $V(z;t)$ satisfies Hypothesis \Sq. Then 
\begin{enumerate}
\item{the function $h(x;t)$ is analytic on $R^1\times [0,t_0]$,}
\item{the free energy $F(t)$ is analytic on $[0,t_0]$,}
\end{enumerate}
\end{cor}

\begin{proof} The analyticity of $h$ follows from formula (\ref{f_e2}) 
and the one of $F$, from (\ref{an:12a}).
\end{proof}

The following extension of Theorem \ref{os_analyticity_q} will be
useful for us. Suppose $V(z;t)$ satisfies Hypothesis \Sq. Then, as
shown in \cite{KM}, for every $t\in(0,t_0]$, there exists $\ep=\ep(t)>0$
such that for any $s\in[1-\ep,\ep]$ the function $s^{-1}V(z;t)$ is
$q$-cut regular and the end-points, $a_i(s;t),b_i(s;t)$ are analytic
in $s\in[1-\ep,\ep]$. 

\begin{prop} 
\label{os_analyticity_q_j}
Suppose $V(z;t)$ satisfies Hypothesis \Sq. Then for any $j\ge 0$ the
functions 
\begin{equation}
\left .\frac{\partial^j a_i(s;t)}{\partial s^j}\right|_{s=1},\quad
\left .\frac{\partial^j b_i(s;t)}{\partial s^j}\right|_{s=1},\qquad
i=1,\dots,q,
\end{equation}
are analytic on $[0,t_0]$.
\end{prop}

\begin{proof} By differentiating system (\ref{an:9}) in $s$ and
  setting $s=1$, we obtain
  a linear $2q\times 2q$ system of equations on 
\begin{equation}\label{an_q_j:1}
\left .\frac{\partial a_i(s;t)}{\partial s}\right|_{s=1},\quad
\left .\frac{\partial b_i(s;t)}{\partial s}\right|_{s=1},\qquad
i=1,\dots,q.
\end{equation}
The  determinant of the system is 
  calculated in (\ref{os:8}), (\ref{os:9}) and it is nonzero. The
  coefficients of the system are analytic function in $t\in[0,t_0]$,
  hence functions (\ref{an_q_j:1})
are analytic in $t\in[0,t_0]$.  By differentiating system (\ref{an:9})
  twice in $s$ and 
  setting $s=1$, we obtain
  a linear $2q\times 2q$ system of equations on 
\begin{equation}\label{an_q_j:2}
\left .\frac{\partial^2 a_i(s;t)}{\partial s^2}\right|_{s=1},\quad
\left .\frac{\partial^2 b_i(s;t)}{\partial s^2}\right|_{s=1},\qquad
i=1,\dots,q.
\end{equation}
The coefficients of the system are the same as for the first
derivatives, but the right hand side changes, and it is expressed in
terms of $a_i,b_i$ and its first derivatives in $s$ at $s=1$, which
are analytic in $t\in[0,t_0]$.
This proves the analyticity of the second derivatives, and so
on. Proposition \ref{os_analyticity_q_j} is proved.
\end{proof}

Let us consider next the coefficients,
$\ga(s;t),\be(s;t),f_{2k}(s;t),g_{2k}(s;t)$, $k\ge 1$, of asymptotic
expansions (\ref{rn:4}) of the recurrence coefficients
for the polynomial $s^{-1}V(z;t)$. 

\begin{prop} 
\label{prop_q_j}
Suppose $V(z;t)$ is a one-parameter family of polynomials of degree
$2d$, which satisfies Hypothesis S. Then for any $j\ge 0$ the
functions 
\begin{equation}\label{gbfg}
\left .\frac{\partial^j \ga(s;t)}{\partial s^j}\right|_{s=1},\quad
\left .\frac{\partial^j \be(s;t)}{\partial s^j}\right|_{s=1},\quad
\left .\frac{\partial^j f_{2k}(s;t)}{\partial s^j}\right|_{s=1},\quad
\left .\frac{\partial^j g_{2k}(s;t)}{\partial s^j}\right|_{s=1},\qquad
k=1,2,\dots,
\end{equation}
are analytic on $[0,t_0]$.
\end{prop}

To prove Proposition \ref{prop_q_j} we will need the following lemma.

\begin{lem}
\label{lemma_q_j}
Suppose $V(z;t)$ is a one-parameter family of polynomials of degree
$2d$,  which satisfies Hypothesis S. Consider  the functions $A(\ga,\be;t)$,
$B(\ga,\be;t)$ corresponding to $V(z;t)$. Then the Jacobian,
\begin{equation}\label{JacAB}
\det
\begin{pmatrix}
\frac{\partial A(\ga,\be;t)}{\partial \ga} &
\frac{\partial A(\ga,\be;t)}{\partial \be} \\
\frac{\partial B(\ga,\be;t)}{\partial \ga} &
\frac{\partial B(\ga,\be;t)}{\partial \be}
\end{pmatrix},
\end{equation}
evaluated at $\ga=\ga(t),\be=\be(t)$,
is analytic and nonzero on $[0,t_0]$.
\end{lem}

\begin{proof}
The analyticity follows from  Theorem \ref{os_analyticity_q}.
Let us prove that the Jacobian is nonzero.
Consider the two-parameter family of polynomials,
\[
V(z;t,t_1)=V(z;t)+t_1z.
\]
Then for every $t\in(0,t_0]$ there exists $\ep=\ep(t)>0$ such 
that $V(z;t,t_1)$ is $q$-cut regular for any $t_1\in[-\ep,\ep]$.
As in Proposition \ref{os_analyticity_q_j}, we obtain that
the functions 
\begin{equation}
\left .\frac{\partial a_i(t_1;t)}{\partial t_1}\right|_{t_1=0},\quad
\left .\frac{\partial b_i(t_1;t)}{\partial t_1}\right|_{t_1=0},\qquad
i=1,\dots,q,
\end{equation}
are analytic on $[0,t_0]$. By using identity (\ref{f_o5}), we obtain
that the Jacobian (\ref{JacAB}) is nonzero.
Lemma \ref{lemma_q_j} is proved.
\end{proof}

\noindent
{\it Proof of Proposition \ref{prop_q_j}.}
The analyticity of $\ga$ and $\be$ follows from (\ref{rec:3}).
To prove the analyticity of  
\begin{equation}\label{sys1}
\left .\frac{\partial \ga(s;t)}{\partial s}\right|_{s=1},\quad
\left .\frac{\partial \be(s;t)}{\partial s}\right|_{s=1},
\end{equation} 
let us differentiate string equations (\ref{jac3}) in $s$ and set
$s=1$. This gives a linear analytic in $t\in[0,t_0]$ system of
equations, whose determinant is nonzero by Lemma \ref{lemma_q_j},
hence functions (\ref{sys1}) are indeed analytic on $[0,t_0]$. By 
differentiating string equations (\ref{jac3}) in $s$ twice we
obtain the analyticity of the second derivatives, and so on. 

Let prove the analyticity of $f_2,g_2$. By following the proof of
Lemmas \ref{first_order}, \ref{any_order} we obtain that the functions
$f_2,g_2$ also satisfy a system of linear equations with the same
coefficients of partial derivatives of $A$ and $B$ and an analytic
right hand side. Hence $f_2,g_2$ are analytic. By differentiating with
respect to $s$ the
system of linear equations on $f_2,g_2$ and setting $s=1$ we obtain a
similar linear system for the derivatives of $f_2,g_2$, and so on. The
same argument applies to $f_4,g_4$ and their derivatives, etc.
Proposition \ref{prop_q_j} is proved.\hfill $\Box$

Now we can prove the one-side analyticity of the coefficients of the
asymptotic expansion of the free energy. We will assume the following
hypothesis. 

{\it Hypothesis T.} $V(z)$ is a polynomial of degree $2d$ such that
\begin{enumerate} 
\item[(a)]{$\tau_t V(z)$ is one-cut regular for $t>1$.}
\item[(b)]{$V(z)$ is one-cut singular and $h(a)\not=0$,
$h(b)\not=0$, where $[a,b]$ is the support of the equilibrium measure
for $V(z)$.}
\end{enumerate}

By Proposition \ref{group}, if $V$ satisfies Hypothesis T,
then $\tau_tV$, $t>1$, satisfies Hypothesis R, hence by Theorem
\ref{expansion_Z},
the free energy $F_N(t)$ of $\tau_tV$ admits the asymptotic expansion,
\begin{equation}\label{os:14}
F_N(t)-F_N^{\Gauss}\sim F(t)+N^{-2}F^{(2)}(t)+N^{-4}F^{(4)}(t)+\dots
\end{equation}

\begin{theo}\label{T1}
Suppose $V(z)$ satisfies Hypothesis T. Then the functions
$F(t)$ and  $F^{(2k)}(t)$, $k\ge 1$, are analytic on $[1,\infty)$. 
\end{theo}
 
\begin{proof} The analyticity of $F(t)$ is proved
in Corollary \ref{cor_os}.
Let us prove the analyticity of $F^{(2j)}(t)$, $j\ge 1$.
To that end substitute expansions (\ref{rn:4}) into (\ref{fin4}), 
and expand the appearing functions $\ga,\be,f_{2k},g_{2k}$ 
in the Taylor series at $n/N=1$. As a result, we obtain asymptotic 
expansion (\ref{os:14}), so that the coefficients $F^{(2j)}(t)$ are
expressed in terms of functions (\ref{gbfg}). By Proposition
\ref{prop_q_j} functions (\ref{gbfg}) are analytic on $[0,t_0]$,
hence the ones $F^{(2j)}(t)$ are analytic as well. 
Theorem \ref{T1} is proved.
\end{proof}

The asymptotics of the partition function for a singular $V$
is a difficult question. The leading term is
defined by the $(N=\infty)$-free energy $F$, see (\ref{an:12}), but 
the subleading terms have a nontrivial scaling. The behavior
of the subleading terms depends on the type of the singular $V$.
The entire problem includes the investigation
of the scaling behavior of the partition
function for a parametric family $V(t)$ passing
through $V$. This is the
problem of the double scaling limit. In the next section we discuss 
the double scaling limit for a singular $V$
of the type I in the terminology of \cite{DKMVZ},
when $h(z)=0$ inside of a cut. We consider a family $V(t)$
of even quartic polynomials passing through the singular polynomial
$V$.

\section{Double Scaling Limit of the Free Energy}\label{DSL}

We will consider the asymptotics of the free energy near the critical
point of the family $\tau_{t}V(z)$ generated by the singular quartic polynomial
$V(z) = \frac{1}{4}\,z^4-z^2$,

\begin{equation}\label{sl:1}
\tau_{t}V(z) \equiv V(z;t) 
=\frac{1}{4t^2}\,z^4+\left(1-\frac{2}{t}\right)z^2 \,.
\end{equation}
We have that $V(z;1)= V(z) = \frac{1}{4}\,z^4-z^2$, and
for $t >1$ the support of the equilibrium measure consists of
one interval, while for $t<1$ it consists of two intervals.
We want to analyse the asymptotics of the free energy
$F_{N}(t)$ as $N\to \infty$ and the parameter $t$ is
confined near its critical value, i.e.  $t=1$. Specifically,
we shall assume the following scaling condition,

$$
|(t-1)N^{2/3}| < C,
$$
and will introduce a
scaling variable $x$ according to the equation
\begin{equation}\label{ds:1}
t=1+N^{-2/3}2^{-2/3}x\,.
\end{equation}
Our aim will be to prove the following theorem
\begin{theo}\label{asymptZx}
Let $F_{N}(t)$ be the partition function corresponding to
the family $V(z;t)$ of quartic potentials (\ref{sl:1}).
Then, for every $\epsilon > 0$,

\begin{equation}\label{Zxas}
F_{N}(t) - F_{N}^{\Gauss} = F^{\reg}_{N}(t) + N^{-2}F^{\sing}_{N}(t)
+ O(N^{-7/3 + \epsilon}), 
\end{equation}
as $N \to \infty$ and $|(t-1)N^{2/3}| < C$. Here,

$$
F^{\reg}_{N}(t) \equiv F(t) + N^{-2} F^{(2)}(t)
$$
is the order $N^{-2}$ (regular at $t=1$) piece of the one-cut expansion (\ref{os:14}),
and

$$
F^{\sing}_{N}(t) = -\log F_{TW}\Big((t-1)2^{2/3}N^{2/3}\Big).
$$
The function $F_{TW}(x)$ is the Tracy-Widom distribution function defined
by the formulae \cite{TW}
\begin{equation}\label{TW}
F_{TW}(x) = \exp \left \{ \int_{x}^{\infty}(x-y)u^{2}(y)dy\right \},
\end{equation}
where $u(y)$ is the Hastings-McLeod solution to the Painlev\'e II 
equation 
\begin{equation}\label{ds:30}
u''(y)=yu(y)+2u^3(y)\,,
\end{equation}
which is characterized by the conditions at infinity \cite{HM},
\begin{equation}\label{ds:40}
\lim_{y\to-\infty}\frac{u(y)}{\sqrt{-\frac{y}{2}}}=1\,,
\qquad \lim_{y\to\infty}\frac{u(y)}{\Ai(y)}=1\,.
\end{equation}
\end{theo}
\begin{proof}
The proof of this theorem is based on the integral representation
(\ref{fin4}) of the free energy wich in  the case of even potentials
$V(z)$  can  be rewriten as follows
\begin{equation}\label{fin4even}
F_{N}(t) = F_{N}^{\Gauss} + \int_{t}^{\infty}\frac{t-\tau}{\tau^2}
\Theta_{N}(\tau)d\tau,
\end{equation}
where
\begin{equation}\label{Theta}
\Theta_{N}(t):=R_{N}(t)(R_{N+1} + R_{N-1}) -\frac{1}{2},
\end{equation}
and we have used a standard notation
\begin{equation}\label{Rn}
\gamma^{2}_{n} \equiv R_{n}.
\end{equation}
Note also that for even potentials all the beta recurrence coefficients
are zero.
Assuming the double scaling substitution (\ref{ds:1}), and making
simultaniously the  change of the variable of integration,
$$
\tau = 1+N^{-2/3}2^{-2/3}y,
$$
we can, in turn, rewrite  (\ref{fin4even}) as
\begin{equation}\label{fin4evenx}
F_{N}(x) = F_{N}^{\Gauss} + 2^{-4/3}N^{-4/3}\int_{x}^{\infty}
\frac{x-y}{(1+N^{-2/3}2^{-2/3}y)^2}
\Theta_{N}(y)dy,
\end{equation}
where,  we use the notations,
\begin{equation}\label{Fx}
F_{N}(x):= F_{N}(t)|_{t =1+N^{-2/3}2^{-2/3}x}
\end{equation}
and
\begin{equation}\label{Thetax}
\Theta_{N}(y):= \Theta_{N}(\tau)|_{\tau =1+N^{-2/3}2^{-2/3}y}.
\end{equation}

Our next move toward the proof of theorem \ref{asymptZx} is to split
the integration in (\ref{fin4evenx}) into the following
two pieces.
$$
 2^{-4/3}N^{-4/3}\int_{x}^{\infty}
\frac{x-y}{(1+N^{-2/3}2^{-2/3}y)^2}
\Theta_{N}(y)dy =  2^{-4/3}N^{-4/3}\int_{x}^{N^{\epsilon}}
\frac{x-y}{(1+N^{-2/3}2^{-2/3}y)^2}
\Theta_{N}(y)dy
$$
\begin{equation}\label{split1}
+ 2^{-4/3}N^{-4/3}\int_{N^{\epsilon}}^{\infty}
\frac{x-y}{(1+N^{-2/3}2^{-2/3}y)^2}
\Theta_{N}(y)dy.
\end{equation}
Going back in the second integral to the original
variable $\tau$, we have the formula,
$$
F_{N}(x) = F_{N}^{\Gauss} +  2^{-4/3}N^{-4/3}\int_{x}^{N^{\epsilon}}
\frac{x-y}{(1+N^{-2/3}2^{-2/3}y)^2}
\Theta_{N}(y)dy
$$
\begin{equation}\label{split2}
+ \int_{1+2^{-2/3}N^{-\delta}}^{\infty}
\frac{1+ N^{-2/3}2^{-2/3}x -\tau}{\tau^2}
\Theta_{N}(\tau)d\tau,
\end{equation}
where
$$
\delta = \frac{2}{3}-\epsilon.
$$
The main point now is that we can produce the uniform
estimates for the recurrence coefficients $R_{n}$,
and hence for the function $\Theta_{N}$, on each
of the two domains of integration. Indeed, the needed
estimates are the extensions to the larger parameter domains
of the  double-scaling asymptotics obtained in \cite{BI2}
(the first integral) and the one-cut asymptotics
obtained (in particular) in \cite{DKMVZ} (the second integral).
Let us first discuss the double-scaling estimates.

Set
\begin{equation}\label{sl:2}
g_{0}=\frac{1}{t^2}\,,\qquad
\kappa=2-\frac{4}{t}\,,
\end{equation}
so that the potential (\ref{sl:1}) is written as
\begin{equation}\label{sl:3}
V(z;t)=\frac{g_{0}}{4}\,z^4+\frac{\kappa}{2}\,z^2 \,.
\end{equation}
Following \cite{BI2}, define $\wh y$ as
\begin{equation}\label{sl:4}
\wh y=c_0^{-1}N^{2/3}\left(\frac{n}{N}-\frac{\kappa^2}{4g_{0}}\right)\,,
\qquad c_0=\left(\frac{\kappa^2}{2g_{0}}\right)^{1/3}\,.
\end{equation}
Then, as shown in \cite{BI2},
\begin{align}\label{sl:5}
R_n(t)&=-\frac{\kappa}{2g_{0}}+N^{-1/3}c_1(-1)^{n+1}u(\wh y)
+N^{-2/3}c_2v(\wh y)+O(N^{-1})\,,\\
\nonumber
c_1&=\left(\frac{2(-\kappa)}{g^2_{0}}\right)^{1/3}\,,\qquad
c_2=\frac{1}{2}\left(\frac{1}{2(-\kappa)g_{0}}\right)^{1/3}\, ,
\end{align}
as $N \to \infty$ and as long as the values of $t$ and $n$ are such that
$\wh{y}$
stays bounded,
\begin{equation}\label{yhatbound}
|\wh y| < C.
\end{equation}
In (\ref{sl:5}), $u(y)$ is the Hastings-McLeod solution to the Painlev\'e II
equation defined in (\ref{ds:30}) - (\ref{ds:40}), and
\begin{equation}\label{ds:5}
v(y)=y+2u^2(y)\,.
\end{equation}
Assume that $t =1+N^{-2/3}2^{-2/3}y$. Then, by simple calculations,
we have
\begin{equation}\label{kappac0}
\frac{\kappa^2}{4g_{0}} = 1 - 2^{1/3}N^{-2/3}y + O(N^{-4/3}y^2),
\quad c_{0}^{-1} = 2^{-1/3} + O(N^{-2/3}y).
\end{equation}
Therefore,
\begin{equation}\label{yhaty}
\wh y = y + O(N^{-2/3}y^2), \quad \mbox{and}\quad
\wh y = y \pm 2^{-1/3} N^{-1/3} + O(N^{-2/3}y^2),
\end{equation}
if $n = N$ and $n = N\pm 1$, respectively. Simultaneously,
\begin{equation}\label{kappa2g}
\frac{\kappa}{2g_{0}} = -1 + O(N^{-4/3}y^2),
\end{equation}
\begin{equation}\label{c12}
c_{1} = 2^{2/3} + O(N^{-2/3}y), \quad
\mbox{and}\quad  c_{2} = 2^{-5/3} + O(N^{-2/3}y).
\end{equation}
 Let (cf. (\ref{Fx}), (\ref{Thetax}))
\begin{equation}\label{Rnx}
 R_{n}(y):= R_{n}(t)|_{t =1+N^{-2/3}2^{-2/3}y},
\end{equation}
and assume that
\begin{equation}\label{ybound}
|y| < C.
\end{equation}
Then, we conclude from (\ref{sl:5}) - (\ref{c12}) that,
as $N\to\infty$, the recurrence coefficients
 $R_n(y)$, $n=N-1,\,N,\,N+1$, have the following
asymptotics:
\begin{align}\label{ds:2}
R_N(y)&=1-N^{-1/3}2^{2/3}(-1)^N u(y)+N^{-2/3}2^{-5/3}v(y)+O(N^{-1})\,,\\
R_{N\pm 1}(y)&=R_N(y)\mp N^{-2/3}2^{1/3}(-1)^N u'(y)+O(N^{-1})\,.
\label{ds:2b}
\end{align}

To be able to use the estimates (\ref{ds:2})-(\ref{ds:2b}) in the
first integral in (\ref{split2}) we need them on the expanding
domain, i.e. we want to be able to replace
the inequality (\ref{ybound}) by the inequality $|y| < N^{\epsilon}$.
\begin{prop}\label{sl}
For every $0 < \epsilon < 1/6$ there
exists a positive constant $ C\equiv C(\epsilon)$
such that the error terms in (\ref{ds:2})-(\ref{ds:2b}),
which we will denote $r_{n}(y)$, $n =N, N+1, N-1$, satisfy the uniform
estimates,
\begin{equation}\label{estdouble}
|r_{n}(y)| \leq C N^{-1 + 2\epsilon}, \quad n = N, N+1, N-1,
\end{equation}
\begin{equation}\label{estdoubl2}
\mbox{for all}\quad  N \geq 1
\quad \mbox{and} \quad |y| < N^{\epsilon}
\end{equation}
\end{prop}

\begin{proof} A simple examination of the proofs of \cite{BI2}
shows that the error term in (\ref{sl:5}) can be specified as
$O(N^{-1}\wh{y}^{3/2})$. This means that, under condition 
$$
|\wh{y}| \leq N^{\epsilon},\quad 0 < \epsilon < \frac{1}{6},
$$
we have from (\ref{sl:5}) the estimate

\begin{equation}\label{sl:5ext}
R_n(t)=-\frac{\kappa}{2g_{0}}+N^{-1/3}c_1(-1)^{n+1}u(\wh y)
+N^{-2/3}c_2v(\wh y)+O(N^{-1 + 3\epsilon/2}).
\end{equation}
Note that the restriction $\epsilon < 1/6$ is needed to 
ensure that the droped terms are of higher order that 
$N^{-2/3}$. Estimate (\ref{sl:5ext}) together with 
(\ref{kappac0} - \ref{c12}) yield the following modification
of (\ref{ds:2}) and (\ref{ds:2b}).

\begin{align}\label{ds:2ext}
R_N(y)&=1-N^{-1/3}2^{2/3}(-1)^N u(y)+N^{-2/3}2^{-5/3}v(y)+O(N^{-1 +2\epsilon})\,,\\
R_{N\pm 1}(y)&=R_N(y)\mp N^{-2/3}2^{1/3}(-1)^N u'(y)+O(N^{-1+ 2\epsilon})\,.
\label{ds:2extb}
\end{align}
The error $O(N^{-1 +2\epsilon})$ is produced by the second term of
(\ref{sl:5ext}). For instance, if $ n = N$, we have

$$
N^{-1/3}c_1u(\wh y)= N^{-1/3}\left(2^{2/3} + O(N^{-2/3}y)\right)
\left(u(y) + O(N^{-2/3}y^2) \right)
$$
$$
= N^{-1/3}u(y) + O(N^{-1}y^2) = N^{-1/3}u(y) + O(N^{-1+2\epsilon}).
$$
Similar arguments lead to (\ref{ds:2extb}). Asymptotics (\ref{ds:2ext})
and (\ref{ds:2extb}) complete the proof of the proposition.
\end{proof}

Let us now turn to the analysis of the one-cut estimates of $R_{n}(t)$
which are needed in the second integral in (\ref{split2}). These estimates
can be
extracted from the general one-cut expansion (\ref{e_Z5a}).
In the case of the quartic potential (\ref{sl:1}), the first two terms
of (\ref{e_Z5a}) can be specified as
\begin{equation}\label{rn:4quar}
R_{n}(t) = R\left(\frac{n}{N};t\right) +
N^{-2}R^{(2)}\left(\frac{n}{N};t\right)
+O(t^{-1}N^{-4}),
\end{equation}
$$
\quad n = N-1,\, N,\, N+1, \quad N\to \infty, \quad t \equiv t_{0} > 1,
$$
where the coefficient functions $R(\lambda;t)$ and
$R^{(2)}(\lambda;t)$ can be found with the help of
the string equation (\ref{rn:3}) which in the case under
consideration takes the form of the single recurrence relation,
\begin{equation}\label{rn:3quartic}
\frac{n}{N} = \left(2-\frac{4}{t}\right) R_{n}
+ \frac{1}{t^2}R_{n}(R_{n+1} + R_{n} + R_{n-1}).
\end{equation}
We also note that the change $t^{-1/2} \to t^{-1}$ in the error estimate 
is due to the evenness of potential (\ref{sl:1}).
Substituting (\ref{rn:4quar}) into equation (\ref{rn:3quartic}) we arrive to
the following explicit formulae for $R(\lambda;t)$ and
$R^{(2)}(\lambda;t)$.
\begin{equation}\label{Rlambda}
R(\lambda;t) = \frac{t}{3}
\left(2-t + \sqrt{(2-t)^2 + 3\lambda}\right),
\end{equation}
\begin{equation}\label{R1lambda}
R^{(2)}(\lambda;t) = -\frac{t}{8}\,
\frac{2-t + \sqrt{(2-t)^2 + 3\lambda}}
{\left((2-t)^2 + 3\lambda\right)^2}.
\end{equation}
We of course need an extension of the validity of the asymptotics
(\ref{rn:4quar}) to the large domain of the parameter $t$.
\begin{prop}\label{onecut}
For every $0 < \delta < 2/3$ there
exists a positive constant $ C\equiv C(\delta)$
such that the error terms in (\ref{rn:4quar}),
which we will denote $r_{n}(t)$, $n =
N, N+1, N-1$, satisfy the uniform estimates,
\begin{equation}\label{est}
|r_{n}(t)| \leq C t^{-1}N^{-4}, \quad n = N, N+1, N-1,
\end{equation}
\begin{equation}\label{est2}
\mbox{for all}\quad  N \geq 1
\quad \mbox{and} \quad t \geq 1 + 2^{-2/3}  N^{-\delta}.
\end{equation}
\end{prop}
The proof of the Proposition is given in Appendix \ref{appendix}.

We are now ready to proceed with the asymptotic evaluation of the
integrals in the right hand side of (\ref{split2}).
We shall start with the first integral,

$$
 2^{-4/3}N^{-4/3}\int_{x}^{N^{\epsilon}}
\frac{x-y}{(1+N^{-2/3}2^{-2/3}y)^2}
\Theta_{N}(y)dy \equiv I_{1}.
$$
Fisrt we notice that,
in virture of Proposition \ref{sl},

\begin{equation}\label{thetaas1}
\Theta_{N}(y) = \frac{3}{2} - 2^{4/3} N^{-2/3}u^{2}(y) +
2^{1/3}N^{-2/3} y + O(N^{-1 + 2\epsilon}).
\end{equation}
Therefore, the $I_{1}$ can be represented as

\begin{equation}\label{firstint1}
I_{1} = 2^{-4/3}N^{-4/3}\int_{x}^{N^{\epsilon}}
\frac{x-y}{(1+N^{-2/3}2^{-2/3}y)^2}
\Theta^{0}_{N}(y)dy
+ O(N^{-7/3 +4\epsilon}).
\end{equation}
where

\begin{equation}\label{thet0}
\Theta^{0}_{N}(y) :=
\frac{3}{2} - 2^{4/3} N^{-2/3}u^{2}(y) +
2^{1/3}N^{-2/3} y.
\end{equation}
Assuming that

\begin{equation}\label{firstint2}
0< \epsilon < \frac{1}{12}\,,
\end{equation}
we make the error term in (\ref{firstint1}) of order $o(N^{-2})$.

The integral in the right hand side of (\ref{firstint1})
can be, in accordance with (\ref{thet0}), splited into the three integrals,
\begin{equation}\label{split333}
 2^{-4/3}N^{-4/3}\int_{x}^{N^{\epsilon}}
\frac{x-y}{(1+N^{-2/3}2^{-2/3}y)^2}
\Theta^{0}_{N}(y)dy \equiv I_{11} + I_{12} + I_{13},
\end{equation}
where

$$
I_{11} =  2^{-4/3}N^{-4/3}\frac{3}{2}\int_{x}^{N^{\epsilon}}
\frac{x-y}{(1+N^{-2/3}2^{-2/3}y)^2}dy,
$$

$$
I_{12} =  - N^{-2}\int_{x}^{N^{\epsilon}}
\frac{x-y}{(1+N^{-2/3}2^{-2/3}y)^2}u^{2}(y)dy,
$$
and

$$
I_{13} =  \frac{1}{2} N^{-2}\int_{x}^{N^{\epsilon}}
\frac{x-y}{(1+N^{-2/3}2^{-2/3}y)^2}ydy,
$$
The integral $I_{11}$ 
can be estimated, up to the terms of order $N^{-2}$, as follows,

$$
I_{11} =  2^{-4/3}N^{-4/3}\frac{3}{2}\int_{x}^{N^{\epsilon}}
(x-y)(1 - 2N^{-2/3}2^{-2/3}y + O(N^{-4/3}y^2))dy
$$
$$
=  2^{-4/3}N^{-4/3}\frac{3}{2}\int_{x}^{N^{\epsilon}}
(x-y)(1 - 2N^{-2/3}2^{-2/3}y)dy + O(N^{-8/3 + 4\epsilon})
$$
$$
= -\frac{3x^2}{2^{10/3}}N^{-4/3} + \frac{x^{3}}{8}N^{-2}
+ \frac{3x}{2^{7/3}}N^{-4/3 + \epsilon} - \frac{3}{2^{10/3}}N^{-4/3 + 2\epsilon}
$$
\begin{equation}\label{I11}
+ \frac{1}{4}N^{-2+3\epsilon} - \frac{3x}{8}N^{-2+2\epsilon} 
+ O(N^{-8/3 + 4\epsilon})
\end{equation}
For the second integral in (\ref{split333}), i.e. for the integral $I_{12}$, we
have

$$
I_{12} =  -N^{-2}\int_{x}^{N^{\epsilon}}
(x-y)(1 + O(N^{-2/3}y))u^{2}(y)dy
$$
$$
=  N^{-2}\int_{x}^{N^{\epsilon}}
(y-x)u^{2}(y)dy + O(N^{-8/3 + 3\epsilon})
$$
\begin{equation}\label{I12}
=  N^{-2}\int_{x}^{\infty}
(y-x)u^{2}(y)dy + O(N^{-8/3 + 3\epsilon})
\end{equation}
and similarly, for the third integral,
$$
I_{13} =  \frac{1}{2}N^{-2}\int_{x}^{N^{\epsilon}}
(x-y)(1 + O(N^{-2/3}y))ydy
$$
$$
= \frac{1}{2} N^{-2}\int_{x}^{N^{\epsilon}}
(x-y)ydy + O(N^{-8/3 + 4\epsilon})
$$
\begin{equation}\label{I13}
=-\frac{x^3}{12}N^{-2} -\frac{1}{6}N^{-2 + 3\epsilon}
+ \frac{x}{4}N^{-2 + 2\epsilon} + O(N^{-8/3 + 4\epsilon})
\end{equation}
Adding the estimates (\ref{I11}), (\ref{I12}), and (\ref{I13}), we
arrive to the following, up to the order $N^{-2}$, asymptotic formula for the first
integral in our basic equation (\ref{split2}),
\begin{equation}\label{I1as1}
I_{1} = -\frac{3x^{2}}{2^{10/3}}N^{-4/3} 
+ \left(\int_{x}^{\infty}
(y-x)u^{2}(y)dy + \frac{x^{3}}{24}\right)N^{-2} + I_{1}(N,\epsilon)
+ O(N^{-7/3 + 4\epsilon}),
\end{equation}
where 
\begin{equation}\label{I1as2}
I_{1}(N,\epsilon): = 
-\frac{3}{2^{10/3}}N^{-4/3 + 2\epsilon}
+ \frac{3x}{2^{7/3}}N^{-4/3+\epsilon} +
\frac{1}{12}N^{-2+3\epsilon} -
\frac{x}{8}N^{-2+2\epsilon}.
\end{equation}

Consider now the second integral in the right hand side of (\ref{split2}),
$$
\int_{1+2^{-2/3}N^{-\delta}}^{\infty}
\frac{1+ N^{-2/3}2^{-2/3}x -\tau}{\tau^2}
\Theta_{N}(\tau)d\tau \equiv I_{2}.
$$
With the help of  Proposition \ref{onecut},
we can specify the general expansion (\ref{FN:1b})
for our case as follows
\begin{equation}\label{thetaas2}
\Theta_{N}(\tau) = \Theta(\tau) + N^{-2}\Theta^{(2)}(\tau) 
+ O(t^{-1}N^{-4}),
\end{equation}
where
\begin{equation}\label{theta0}
\Theta(\tau) = 2R^{2}(1;\tau) - \frac{1}{2},
\end{equation}
and
\begin{equation}\label{theta2}
\Theta^{(2)}(\tau) = 4R(1;\tau)R^{(2)}(1;\tau)
+R(1;\tau)R_{\lambda \lambda}(1;\tau).
\end{equation}
Therefore, similar to the integral $I_{1}$, we can
represent the integral $I_{2}$, up to the terms of
order $N^{-2}$, as the sum of the following three integrals,
\begin{equation}\label{I2as1}
I_{2} = I_{21} + I_{22} + I_{23} + O(N^{-8/3}),
\end{equation}
where
\begin{equation}\label{I21}
I_{21} = \int_{1+2^{-2/3}N^{-\delta}}^{\infty}
\frac{1 -\tau}{\tau^2}
\Theta(\tau)d\tau,
\end{equation}
\begin{equation}\label{I22}
I_{22} = N^{-2/3}2^{-2/3}x\int_{1+2^{-2/3}N^{-\delta}}^{\infty}
\frac{1}{\tau^2}
\Theta(\tau)d\tau,
\end{equation} 
and
\begin{equation}\label{I23}
I_{23} = N^{-2}\int_{1+2^{-2/3}N^{-\delta}}^{\infty}
\frac{1 -\tau}{\tau^2}
\Theta^{(2)}(\tau)d\tau.
\end{equation}

In analysing each of the integrals $I_{2k}$ we shall recall
that
$$
\delta = \frac{2}{3} - \epsilon,
$$
and make use of the following elementary estimate,
\begin{equation}\label{taylorest}
\int_{1+s}^{\infty}f(\tau)d\tau = 
\int_{1}^{\infty}f(\tau)d\tau - f(1)s -\frac{s^2}{2}f'(1) -\frac{s^3}{6}f''(1)
+ O(s^4)
\end{equation}
which is true under the natural conditions fulfiled in the case
of each of the integrals $I_{2k}$. Applying (\ref{taylorest}) to the 
integral $I_{21}$ we obtain the asymptotic relation,
$$
I_{21} = \int_{1}^{\infty}\frac{1 -\tau}{\tau^2}\Theta(\tau)d\tau
-\frac{1}{2}\left(\frac{1 -\tau}{\tau^2}\Theta(\tau)\right)'\Big|_{\tau =1}
2^{-4/3}N^{-4/3 + 2\epsilon}
$$
\begin{equation}\label{I21as1}
-\frac{1}{6}\left(\frac{1 -\tau}{\tau^2}\Theta(\tau)\right)''\Big|_{\tau =1}
2^{-2}N^{-2 + 3\epsilon}
+O(N^{-8/3 +4\epsilon})
\end{equation}
Observe that
$$
\left(\frac{1 -\tau}{\tau^2}\Theta(\tau)\right)'\Big|_{\tau =1}
= - \Theta(1),
$$
and
$$
\left(\frac{1 -\tau}{\tau^2}\Theta(\tau)\right)''\Big|_{\tau =1}
= 4\Theta(1) -2\Theta'(1).
$$
This, together with equations (\ref{theta0}) 
and (\ref{Rlambda}) allows us to evaluate the coefficients
of expansion (\ref{I21as1}). Indeed we have
\begin{equation}\label{thetader1}
\Theta(1) = \frac{3}{2}, \quad \Theta'(1) = 2,
\end{equation}
and hence

$$
I_{21} = \int_{1}^{\infty}\frac{1 -\tau}{\tau^2}\Theta(\tau)d\tau
+ \frac{3}{2^{10/3}}N^{-4/3 + 2\epsilon}
$$
\begin{equation}\label{I21as2}
-\frac{1}{12}N^{-2 + 3\epsilon}
+O(N^{-8/3 +4\epsilon}).
\end{equation}

Similarly, for the integral $I_{22}$ we have
$$
I_{22} = N^{-2/3}2^{-2/3}x\int_{1}^{\infty}
\frac{1}{\tau^2}
\Theta(\tau)d\tau 
$$
$$
-  x\Theta(1)2^{-4/3}N^{-4/3 + \epsilon}
-  \frac{x}{2}\left(\frac{1}{\tau^2}\Theta(\tau)\right)'\Big|_{\tau =1}
2^{-2}N^{-2 + 2\epsilon}
+ O(N^{-8/3 + 3\epsilon})
$$

$$
= N^{-2/3}2^{-2/3}x\int_{1}^{\infty}
\frac{1}{\tau^2}
\Theta(\tau)d\tau 
$$
$$
-  x\Theta(1)2^{-4/3}N^{-4/3 + \epsilon}
+  \frac{x}{8}\left(2\Theta(1) - \Theta'(1)\right)
N^{-2 + 2\epsilon}
+ O(N^{-8/3 + 3\epsilon}),
$$
and, taking into account (\ref{thetader1}),

$$
I_{22} = N^{-2/3}2^{-2/3}x\int_{1}^{\infty}
\frac{1}{\tau^2}
\Theta(\tau)d\tau
$$
\begin{equation}\label{I22as1}
 -  \frac{3x}{2^{7/3}}N^{-4/3 +
\epsilon} +  \frac{x}{8}
N^{-2 + 2\epsilon}
+ O(N^{-8/3 + 3\epsilon})
\end{equation} 

The estimation of the integral $I_{23}$ up to the
order $N^{-2}$ is very simple - we only  need to use the first
term of (\ref{taylorest}):

\begin{equation}\label{I23as1}
I_{23} = N^{-2}\int_{1}^{\infty}
\frac{1-\tau}{\tau^2}
\Theta^{(2)}(\tau)d\tau
+ O(N^{-8/3 + \epsilon})
\end{equation}
 
Adding the estimates (\ref{I21as2}), (\ref{I22as1}) and
(\ref{I23as1}) we conclude that

$$
I_{2} =  \int_{1}^{\infty}\frac{1 -\tau}{\tau^2}\Theta(\tau)d\tau
+ N^{-2/3}2^{-2/3}x\int_{1}^{\infty}
\frac{1}{\tau^2}
\Theta(\tau)d\tau
+ N^{-2}\int_{1}^{\infty}
\frac{1-\tau}{\tau^2}
\Theta^{(2)}(\tau)d\tau
$$
\begin{equation}\label{I2as2}
-I_{1}(N,\epsilon) +  O(N^{-8/3 + 4\epsilon}),
\end{equation}
where $I_{1}(N,\epsilon)$ is exactly the same collection
of the epsilon-depending terms as the one which has appeared in formula
(\ref{I1as1}) evaluating the integral $I_{1}$, and which is defined in
(\ref{I1as2}).

Substituting estimates (\ref{I1as1}) and (\ref{I2as2}) into the
basic equation (\ref{split2}) we obtain the following asymptotic
representation of the free energy $F_{N}(x)$, 

$$
F_{N}(x) = F_{N}^{\Gauss} + \int_{1}^{\infty}\frac{1
-\tau}{\tau^2}\Theta(\tau)d\tau
+ N^{-2/3}2^{-2/3}x\int_{1}^{\infty}
\frac{1}{\tau^2}
\Theta(\tau)d\tau
$$

$$
-\frac{3x^{2}}{2^{10/3}}N^{-4/3} 
+ N^{-2}\left(\int_{x}^{\infty}
(y-x)u^{2}(y)dy + \frac{x^{3}}{24}
+\int_{1}^{\infty}
\frac{1-\tau}{\tau^2}
\Theta^{(2)}(\tau)d\tau \right)
$$

\begin{equation}\label{freedsas}
+ O(N^{-7/3 + 4\epsilon})
\end{equation}
Put (cf. (\ref{os:14}))

$$
F_{N}^{\reg}(t)\equiv F(t) + N^{-2}F^{(2)}(t)
$$

\begin{equation}\label{Freg}
= \int_{t}^{\infty}\frac{t
-\tau}{\tau^2}\Theta(\tau)d\tau +
N^{-2}\int_{t}^{\infty}\frac{t
-\tau}{\tau^2}\Theta^{(2)}(\tau)d\tau,
\end{equation}
and consider

$$
F_{N}^{\reg}(1 + 2^{-2/3}N^{-2/3}x).
$$
It is easy to see that this object coinside with the
sum $I_{21} + I_{22} + I_{23}$ (see (\ref{I2as1})) up
to  the following formal replaicment:
$$
N^{\epsilon} \to x.
$$
Therefore, we can apply (\ref{I2as2}) and see that

$$
F_{N}^{\reg}(1 + 2^{-2/3}N^{-2/3}x)
= \int_{1}^{\infty}\frac{1
-\tau}{\tau^2}\Theta(\tau)d\tau
+ N^{-2/3}2^{-2/3}x\int_{1}^{\infty}
\frac{1}{\tau^2}
\Theta(\tau)d\tau
$$

$$
+ N^{-2}\int_{1}^{\infty}
\frac{1-\tau}{\tau^2}
\Theta^{(2)}(\tau)d\tau 
-\frac{3x^{2}}{2^{10/3}}N^{-4/3} +  \frac{x^{3}}{24}
$$
\begin{equation}\label{freereg}
+ O(N^{-8/3})
\end{equation}

This allows us to rewrite the final equation (\ref{freedsas})
as

$$
F_{N}(x) = F_{N}^{\Gauss} + F_{N}^{\reg}(1 + 2^{-2/3}N^{-2/3}x)
$$

\begin{equation}\label{freedsas2}
-N^{-2}\log F_{TW}(x) + O(N^{-7/3 +4\epsilon}),
\end{equation}
which concludes the proof of theorem \ref{asymptZx}.
\end{proof}

{\it Remark.} In terms of the  partition function 
equation (\ref{freedsas2}) reads

$$
\frac{Z_{N}(t)}{Z_{N}^{\Gauss}}
=F_{TW}\Big((t-1)2^{2/3}N^{2/3}\Big)Z_{N}^{\reg}(t)
$$
\begin{equation}\label{freedsas3}
\times \Big(1 + O(N^{-1/3 +\epsilon})\Big),
\end{equation}
where $\epsilon$ is an arbitrary positive number.

\appendix
\section{The proof of proposition \ref{onecut}}\label{appendix}

Let us remind the basic steps of
the Riemann-Hilbert approach to the asymptotic analysis of  orthogonal polynomial
following the scheme of \cite{DKMVZ}.

The principal observation (\cite{FIK}; see also \cite{BI1} and \cite{DKMVZ}) is that
the orthogonal polynomials $P_{n}(z)$ admit the representation,
\begin{equation}\label{polY}
P_{n}(z) = Y_{n11}(z),
\end{equation}
where the $2\times 2$ matrix function $Y_{n}(z)$ is the (unique) solution of the following
Riemann-Hilbert (RH) problem.
\begin{enumerate}
\item $Y(z)$ is analytic for $z \in {\mathbb C} \setminus {\mathbb R}$,
and it has continuous limits, $Y_{n+}(z)$ and $Y_{n-}(z)$ from
above and below the real line,
$$
Y_{n\pm}(z) = \lim_{z' \to z,\,\, \pm \,\mbox{Im}\,\, z'\, >0}Y_{n}(z').
$$
\item $Y_{n}(z)$ satisfies the jump condition on the real line,
\begin{equation}\label{Yjump}
Y_{n+}(z) = Y_{n-}(z)G(z),
\end{equation}
where
\begin{equation}\label{jumpmat}
G(z) = \begin{pmatrix}
1 & e^{-NV(z)} \\
0 & 1
\end{pmatrix}
\end{equation}
\item as $z \to \infty$, the function $Y_{n}(z)$ has the following uniform
asymptotics expansion:
\begin{equation}\label{Yasymp}
Y_{n}(z) \sim \left(I + \sum_{k=1}^{\infty}\frac{m^{(n)}_{k}}{z^k}\right)
z^{n\sigma_{3}}, \quad z \to \infty,
\end{equation}
where
$$
\sigma_{3} = \begin{pmatrix}
1 & 0\\
0 &-1
\end{pmatrix}.
$$
\end{enumerate}
In addition to equation (\ref{polY}), the recurrence coefficients $R_{n}$
can be also evaluated directly via $Y_{n}(z)$. In fact, we have that
\begin{equation}\label{coefY}
R_{n} = (m^{(n)}_{1})_{12}(m^{(n)}_{1})_{21},
\end{equation}
where the matrix $m^{(n)}_{1}$ is the first coefficient  of the asymptotic
series (\ref{Yasymp}). Equation (\ref{coefY}) reduces the question of the
asymptotic investigation of the recurrence coefficients $R_{n}(t)$ to
the question of the asymptotic solution of the RH problem (1-3).
In the case of a fixed $t > 1$,  this analysis
is performed  in \cite{DKMVZ}. In fact, in \cite{DKMVZ} 
the asymptotics is evaluate for a generic fixed real analytic potential $V(z)$.
The approach of \cite{DKMVZ} consists of a succession of steps which, 
in the end, yields a reduced RH problem in which all the jumps  are of the order
$N^{-1}$ (cf. the proof of theorem \ref{asympt}).  
In the relevant for our analysis one-cut situation, these steps
are described in detail in \cite{EM}. In what follows we will repeat
the construction of \cite{EM} specifying its principal ingredients for
the case of the potential (\ref{sl:1}) and  showing how
it can be modified in order to cover the extanded range of parameter $t$, i.e.
assuming $t
\geq 1 + 2^{-2/3}N^{-\delta}$.

{\it Step1.} ({\it $g$-function deformation}). Define
\begin{equation}\label{gdef}
g(z) := \int_{-z_{0}}^{z_{0}}\ln(z-s)\rho(s)ds,
\end{equation}
where $[-z_{0}, z_{0}]$ and $\rho(s)$ are the support
and the density of the equilibrium measure (\ref{rn:2}),
respectively. More precisely, $[-z_{0}, z_{0}]$ and $\rho(s)$
minimise the functional (\ref{an:2}) where $V(x)$ is replaced by
$\frac{1}{\lambda}V(x) $, and $\lambda = n/N$, $n = N-1, N, N+1$. 
In the case of the potential (\ref{sl:1}),  the point $z_{0}$ and
the function $\rho(s)$ are given by the equations (see, e.g. \cite{BPS}),
\begin{equation}\label{z0}
z_{0} = 2 \left(\frac{t}{3}\left(2-t + 
\sqrt{(t-2)^2 +3\lambda}\right)\right)^{\frac{1}{2}}
\equiv 2 R^{1/2}(\lambda;t),
\end{equation}
\begin{equation}\label{rhods}
\rho(s) = \frac{1}{\pi \lambda}(b_{0} + b_{2}s^{2})\sqrt{z^2_{0} - s^{2}}
\equiv \frac{1}{\pi i \lambda}(b_{0} + b_{2}s^{2})\left(\sqrt{s^2 - z_{0}^{2}}\right)_{+},
\end{equation}
where 
\begin{equation}\label{b0}
b_{0} = \frac{1}{3t}
\left(2t-4 + \sqrt{(t-2)^2 +3\lambda}\right),
\end{equation}
and
\begin{equation}\label{b2}
b_{2} = \frac{1}{2t^2}.
\end{equation}
The branch of $\sqrt{z^2 - z_{0}^{2}}$ is defined on ${\mathbb C}\setminus
[-z_{0}, z_{0}]$ and is fixed by the condition
$\sqrt{z^2 - z_{0}^{2}} > 0$ if $z > z_{0}$. The branch of 
$\ln(z-s)$ is defined on ${\mathbb C}\setminus
(-\infty, s]$ and is fixed by the condition
$\arg(z-s) = 0$ if $z > s$. 

Assume that $t \geq t_{0} > 1$ and denote,
$$
V_{\lambda}(x) \equiv \frac{1}{\lambda}V(x).
$$
Then the function $g(z)$ satisfies 
the following characteristic properties (cf. (\ref{an:6a})-
(\ref{an:6b})) which underline the importance of $g(z)$
for the asymptotic analysis of the RH problem
(1-3).
\begin{itemize}
\item  The function $g(z)$ is analytic for $z \in {\mathbb C}\setminus
(-\infty, z_{0}]$ with continuous boundary values $g_{\pm}(z)$
on $(-\infty, z_{0}]$.
\vskip.2in
\item There is a constant $l$ such that  for $z \in [-z_{0}, z_{0}]$,

\begin{equation}\label{jumpg1}
g_{+}(z) + g_{-}(z) - V_{\lambda}(z) = l,
\end{equation}
and for $z \in {\mathbb R} \setminus [-z_{0}, z_{0}]$,

\begin{equation}\label{jumpg2}
g_{+}(z) + g_{-}(z) - V_{\lambda}(z) < l.
\end{equation}
\vskip.2in
\item Denote
\begin{equation}\label{pdef}
p(z):=g_{+}(z) - g_{-}(z).
\end{equation}
Then, for $z \in [-z_{0}, z_{0}]$,

\begin{equation}\label{jumpg3}
p(z) = 2\pi i\int_{z}^{z_{0}}\rho(s)ds,
\end{equation}
and this function possesses an analytic continuation to a neighborhood of
$(-z_{0}, z_{0})$. Moreover, for every $0<d < z_{0}/2$ there is 
a positive number $p_{0}$ such that

\begin{equation}\label{pderiv}
\frac{d}{d\sigma}\,\mbox{Re}\,p(s+i\sigma)\Big|_{\sigma = 0}
=\frac{2}{\lambda}(b_{0} + b_{2}s^{2})\sqrt{z^2_{0} - s^{2}} \geq p_{0} >0,
\end{equation}
\vskip.2in
\noindent
for all $s\in [-z_{0}+d, z_{0}-d ]$ and $t \geq t_{0} > 1$.
\vskip.2in
\item For $z > z_{0}$,

\begin{equation}\label{jumpg4}
p(z) = 0,
\end{equation}
and for $z < -z_{0}$,

\begin{equation}\label{jumpg5}
p(z) = 2\pi i.
\end{equation}
\vskip.2in
\item as $z \to \infty$,  
\begin{equation}\label{gnorm}
g(z) = \ln z + O\left(\frac{1}{z^2}\right).
\end{equation}
\end{itemize}
We also notice that there is the following alternative representation
of the function $g(z)$,

\begin{equation}\label{galt}
g(z) = -\frac{1}{\lambda}
\int_{z_{0}}^{z}(b_{0} + b_{2}s^{2})\sqrt{s^2 - z_{0}^{2}}\,ds
 +\frac{1}{2}V_{\lambda}(z) + \frac{l}{2}.
\end{equation}

Having introduced the function $g(z)$ and the constant $l$,
we define the  first transformation, $Y(z) \to \Phi(z)$ of 
the original RH problem, by the equation,

\begin{equation}\label{first} 
Y(z) = e^{\frac{nl}{2}\sigma_{3}}
\Phi(z)e^{n(g(z) - \frac{l}{2})\sigma_{3}}.
\end{equation} 
In terms of the function $\Phi(z)$ the RH problem 
(1-3) reads as follows.
\vskip .2in
\noindent
(1$'$)\quad $\Phi(z)$ is analytic for $z \in {\mathbb C} \setminus {\mathbb R}$.
\vskip .2in
\noindent
(2$'$)\quad $\Phi(z)$ satisfies the jump condition on the real line,
\begin{equation}\label{Phijump}
\Phi_{+}(z) = \Phi_{-}(z)G_{\Phi}(z),
\end{equation}
where
\begin{equation}\label{Phijumpmat}
G_{\Phi}(z) = \begin{pmatrix}
e^{-np(z)} & e^{n\left(g_{+}(z) + g_{-}(z) - V_{\lambda} - l\right)} \\
0 & e^{np(z)}
\end{pmatrix}
\end{equation}
\vskip .2in
\noindent
(3$'$)\quad as $z \to \infty$, the function $\Phi(z)$ has the following uniform
asymptotics:
\begin{equation}\label{Phiasymp}
\Phi(z) = I + 0\left(\frac{1}{z}\right),
\quad z \to \infty
\end{equation}
(which can be extended to the whole asymptotic series).
 
Observe that, in virtue of (\ref{jumpg4}) and (\ref{jumpg5}),
we have 

\begin{equation}\label{GPhi1}
G_{\Phi}(z) = \begin{pmatrix}
1 & e^{n\left(g_{+}(z) + g_{-}(z) - V_{\lambda} - l\right)} \\
0 & 1
\end{pmatrix}, \quad \mbox{for}\quad z \in {\mathbb R}\setminus [-z_{0}, z_{0}],
\end{equation}
and in virtue of (\ref{jumpg1}),

\begin{equation}\label{GPhi2}
G_{\Phi}(z) = \begin{pmatrix}
e^{-np(z)} & 1 \\
0 & e^{np(z)}
\end{pmatrix}, \quad \mbox{for}\quad z \in [-z_{0}, z_{0}].
\end{equation}
\
{\it Step2.} ({\it Second transformation $\Phi \to \Phi^{(1)}$})
Next we introduce the lens-shaped region $\Omega = \Omega^{(u)}
\cup \Omega^{(l)}$ around $(-z_{0}, z_{0})$ as 
indicated in Figure 1 and define $\Phi^{(1)}(z)$ as follows

\begin{figure}
\scalebox{0.65}{\includegraphics{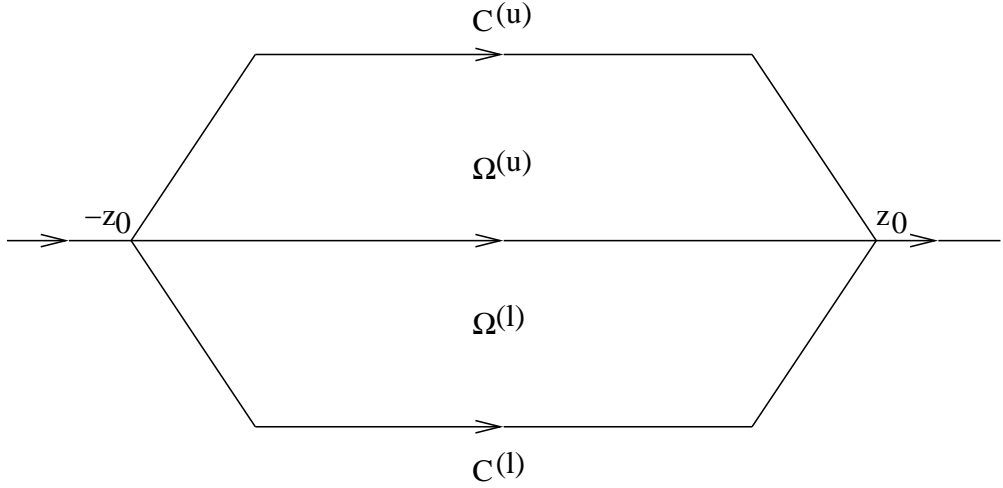}}
\caption{The contour $\Ga$.}
\label{fig1}
\end{figure}

\vskip .2in
\noindent
(i)\quad for $z$ outside the domain $\Omega$,

\begin{equation}\label{Phi11}
\Phi^{(1)}(z) = \Phi(z);
\end{equation}
\vskip .2in
\noindent
(ii)\quad for $z$ within the domain $\Omega^{(u)}$ (the upper lens),

\begin{equation}\label{Phi12}
\Phi^{(1)}(z) = \Phi(z)\begin{pmatrix}
1 & 0 \\
-e^{-np(z)} & 1
\end{pmatrix};
\end{equation}
\vskip .2in
\noindent
(ii)\quad for $z$ within the domain $\Omega^{(l)}$ (the lower lens),

\begin{equation}\label{Phi13}
\Phi^{(1)}(z) = \Phi(z)\begin{pmatrix}
1 & 0 \\
e^{np(z)} & 1
\end{pmatrix};
\end{equation}
(We note that the function $p(z)$ admits the analytic continuation to
the domain $\Omega$.)

With the passing to $\Phi^{(1)}(z)$, the RH problem
(1$'$ - 3$'$) transforms to the RH problem posed
on the contour $\Gamma$ consisting of the real axes 
and the curves $C^{(u)}$ and $C^{(l)}$ which
form the boundary of the domain $\Omega$,
$$
\Omega = C^{(l)} - C^{(u)}
$$
(see Figure 1). We have,

\vskip .2in
\noindent
(1$''$)\quad $\Phi^{(1)}(z)$ is analytic for $z \in {\mathbb C} \setminus {\Gamma}$.
\vskip .2in
\noindent
(2$''$)\quad $\Phi^{(1)}(z)$ satisfies the jump condition on the real line,
\begin{equation}\label{Phi1jump}
\Phi^{(1)}_{+}(z) = \Phi^{(1)}_{-}(z)G_{\Phi^{(1)}}(z),
\end{equation}
where
\begin{equation}\label{Phi1jumpmat}
G_{\Phi^{(1)}}(z) = \left\{\begin{array}{ll}
\begin{pmatrix}
1 & e^{n\left(g_{+}(z) + g_{-}(z) - V_{\lambda} - l\right)} \\
0 & 1
\end{pmatrix},& \mbox{for} \, \, 
z \in {\mathbb R}\setminus [-z_{0}, z_{0}],\\\\
\begin{pmatrix}
1 & 0 \\
e^{-np(z)} & 1
\end{pmatrix},& \mbox{for} \, \, z \in C^{(u)},\\\\
\begin{pmatrix}
1 & 0 \\
e^{np(z)} & 1
\end{pmatrix},& \mbox{for} \, \, z \in C^{(l)},\\\\
\begin{pmatrix}
0 & 1 \\
-1 & 0
\end{pmatrix},& \mbox{for} \, \, z \in [-z_{0}, z_{0}]
\end{array}\right.
\end{equation}
\vskip .2in
\noindent
(3$''$)\quad as $z \to \infty$, the function $\Phi^{(1)}(z)$ has 
the following uniform
asymptotics:
\begin{equation}\label{Phi1asymp}
\Phi^{(1)}(z) = I + 0\left(\frac{1}{z}\right),
\quad z \to \infty
\end{equation}
(which can be extended to the whole asymptotic series).
Indeed, in view of the equations (\ref{Phi11}) - (\ref{Phi13})
defining the function $\Phi^{(1)}(z)$, the properties (1$'$) - (3$'$)
of the function $\Phi(z)$ and equation (\ref{GPhi1}),
we only need to explane the last line of equation (\ref{Phi1jumpmat}).
The latter is a direct consequance of equation (\ref{GPhi2}) and
the elementary algebraic identity,

$$
\begin{pmatrix}
0 & 1 \\
-1 & 0
\end{pmatrix} =
\begin{pmatrix}
1 & 0 \\
-e^{np} & 1
\end{pmatrix}
\begin{pmatrix}
e^{-np} & 1 \\
0 & e^{np}
\end{pmatrix}
\begin{pmatrix}
1 & 0 \\
-e^{-np} & 1
\end{pmatrix}
$$

{\it Step3.} ({\it The construction of a global aproximation to $\Phi^{(1)}(z)$})
The point of the transformation of the original $Y$ - RH problem (1 - 3) to 
the $\Phi$ - RH problem (1$''$ - 3$''$) is that in virtue of the
inequalities (\ref{jumpg2}) and (\ref{pderiv}), the jump matrix 
$G_{\Phi^{(1)}}(z)$, for  $z \neq \pm z_{0}$, is exponentially
close to the identity matrix on the part $\Gamma \setminus [-z_{0}, z_{0}]$
of the jump contour $\Gamma$, so that one can expect that,
as $N \to \infty$, $n = N-1, N, N+1$, and $|z\pm z_{0}| > \delta$,

\begin{equation}\label{Phiinf1}
\Phi^{(1)}(z) \sim \Phi^{(\infty)}(z),\
\end{equation}
where $\Phi^{(\infty)}(z)$ is the solution of the following
model RH problem.

\vskip .2in
\noindent
(1$'''$)\quad $\Phi^{(\infty)}(z)$ is analytic for $z \in {\mathbb C} \setminus 
[-z_{0}, z_{0}]$.
\vskip .2in
\noindent
(2$'''$)\quad $\Phi^{(\infty)}(z)$ satisfies the jump condition on $(-z_{0}, z_{0})$
\begin{equation}\label{Phiinfjump}
\Phi^{(\infty)}_{+}(z) = \Phi^{(\infty)}_{-}(z)
\begin{pmatrix}
0 & 1 \\
-1 & 0
\end{pmatrix}.
\end{equation}
\vskip .2in
\noindent
(3$'''$)\quad as $z \to \infty$, the function $\Phi^{(\infty)}(z)$ 
has the following uniform asymptotics:
\begin{equation}\label{Phiinfasymp}
\Phi^{(\infty)}(z) = I + 0\left(\frac{1}{z}\right),
\quad z \to \infty
\end{equation}
(which can be extended to the convergent Laurent series at $z = \infty$).
The important fact is that this Riemann-Hilbert problem admits an explicit solution:

\begin{equation}\label{Phiinf2}
\Phi^{(\infty)}(z)
= \begin{pmatrix}
\frac{\alpha + \alpha^{-1}}{2} & \frac{\alpha - \alpha^{-1}}{2i}  \\
\frac{\alpha - \alpha^{-1}}{-2i} & \frac{\alpha + \alpha^{-1}}{2}
\end{pmatrix},
\end{equation}

\begin{equation}\label{Phiinf3}
\alpha(z) = \left(\frac{z-z_{0}}{z+z_{0}}\right)^{1/4},
\quad  \alpha(\infty) = 1.
\end{equation}

In order to prove and specify the error term in estimation (\ref{Phiinf1})
we need to construct the parametrix of the solution $\Psi^{(1)}(z)$
near the end points $\pm z_{0}$. Let $B_{d}$ denote a disc
of radius $d$ centered at $z_{0}$, and let us introduce
the change-of-the-variable function $w(z)$ on $B_{d}$
by the formula,

\begin{equation}\label{w}
w(z) = \left(\frac{3}{4}\right)^{2/3}(-2g(z) +V_{\lambda}(z) + l)^{2/3}.
\end{equation}
In  view of equation (\ref{galt}), the function $w(z)$
can be also written as,

\begin{equation}\label{walt}
w(z) = \left(\frac{3}{4}\right)^{2/3}
\left(\frac{2}{\lambda}\int_{z_{0}}^{z}(b_{0} + b_{2}s^{2})
\sqrt{s^{2} - z^{2}_{0}}\, ds\right)^{2/3},
\end{equation}
which, taking into account that $|b_{0} + b_{2}z_{0}^{2}|> c_{0} >0$
for all $t \geq 1$, implies that, for sufficiently small $d$,
the function $w(z)$ is holomorphic and in fact conformal in the disc 
$B_{d}$,

\begin{equation}\label{wser}
w(z) = \sum_{k=1}^{\infty}w_{k}(z-z_{0})^{k}, \quad z \in B_{d}.
\end{equation} 
We shall assume that the branch of the root $(\,)^{2/3}$ is
choosen in such a way that 

\begin{equation}\label{branch}
w_{1}\geq c_{0} > 0 \quad \mbox{for all}\quad t\geq 1 
\quad \mbox{and} \quad N \geq 1.
\end{equation} 
We also note that, for sufficiently small $d$,
the following inequality takes place,

\begin{equation}\label{wineq}
|w(z)| \geq c_{0}, \quad \mbox{for all}\quad z \in S_{d},
\quad t\geq 1\quad \mbox{and} \quad  N \geq 1,
\end{equation} 
where $S_{d}$ denote the boundary of $B_{d}$,
i.e the circle of radius $d$ centered at $z_{0}$.

Let us decompose $B_{d}$ into four regions (see Figure 2),

\begin{figure}
\scalebox{0.7}{\includegraphics{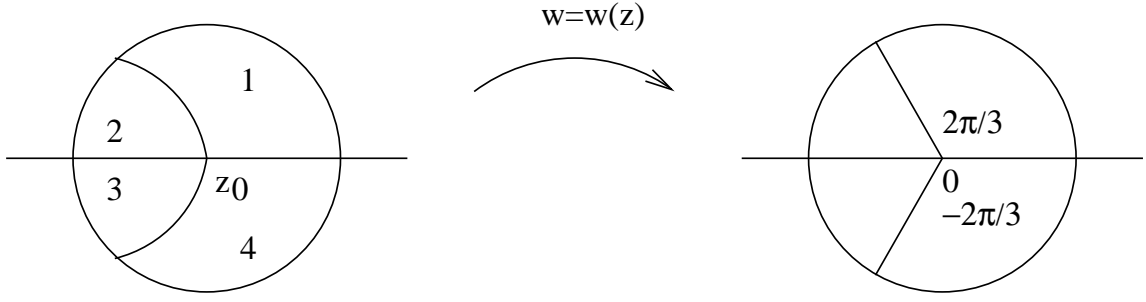}}
\caption{Decomposition of $B_d$.}
\label{fig2}
\end{figure}

\begin{equation}\label{bdecomp}
B_{d} = B_{d}^{(1)}\cup B_{d}^{(2)}
\cup B_{d}^{(3)}\cup  B_{d}^{(4)},
\end{equation}
where 

$$
B_{d}^{(1)} = \left \{ z \in B_{d}: 
0 \leq \arg w(z) \leq \frac{2\pi}{3} \right \},
$$

$$
B_{d}^{(2)} = \left \{ z \in B_{d}: 
\frac{2\pi}{3} \leq \arg w(z) \leq \pi \right \},
$$

$$
B_{d}^{(3)} = \left \{ z \in B_{d}: 
-\pi \leq \arg w(z) \leq -\frac{2\pi}{3} \right \},
$$

$$
B_{d}^{(4)} = \left \{ z \in B_{d}: 
-\frac{2\pi}{3} \leq \arg w(z) \leq 0 \right \}.
$$
We shall assume that the parts of the curves $C^{(u,l)}$
which are inside $B_{d}$ coincide with the relevant
partys of the boundaries of the domains $B^{(k)}_{d}$.
Let us also introduce the standard 
collection of the Airy functions,
 
\begin{equation}\label{airy}
y_{0}(z):= \mbox{Ai}\,(z), \quad
y_{1}(z):= e^{-{\pi i}/6}\mbox{Ai}\,(e^{-2\pi i/3}z)\quad
y_{2}(z):= e^{{\pi i}/6}\mbox{Ai}\,(e^{2\pi i/3}z)
\end{equation}
We will now define the approximation
(parametrix) $\Phi^{(z_{0})}(z)$  within $B_{\delta}$ by the
following equation,

\begin{equation}\label{Phiend1}
\Phi^{(z_{0})}(z) = E(z)n^{\frac{1}{6}\sigma_{3}}
\left\{\begin{array}{ll}
\Psi^{(u)}_{\mbox{Ai}}(n^{2/3}w(z))e^{\frac{2n}{3}w^{3/2}(z)
\sigma_{3}},& \mbox{for} \, \, 
z \in B^{(1)}_{d},\\\\
\Psi^{(u)}_{\mbox{Ai}}(n^{2/3}w(z))
\begin{pmatrix}
1 & 0 \\
-1 & 1
\end{pmatrix}
e^{\frac{2n}{3}w^{3/2}(z)
\sigma_{3}},& \mbox{for} \, \, 
z \in B^{(2)}_{d},\\\\
\Psi^{(l)}_{\mbox{Ai}}(n^{2/3}w(z))
\begin{pmatrix}
1 & 0 \\
1 & 1
\end{pmatrix}
e^{\frac{2n}{3}w^{3/2}(z)
\sigma_{3}},& \mbox{for} \, \, 
z \in B^{(3)}_{d}, \\\\
\Psi^{(l)}_{\mbox{Ai}}(n^{2/3}w(z))e^{\frac{2n}{3}w^{3/2}(z)
\sigma_{3}},& \mbox{for} \, \, 
z \in B^{(4)}_{d}
\end{array}\right.
\end{equation}
where the model functions  $\Psi^{(u, d)}_{\mbox{Ai}}(z)$
are the matrices,

\begin{equation}\label{PsiAiu}
\Psi^{(u)}_{\mbox{Ai}}(z)
=\begin{pmatrix}
y_{0}(z) & iy_{1}(z) \\
y'_{0}(z) & iy'_{1}(z)
\end{pmatrix},
\end{equation}

\begin{equation}\label{PsiAid}
\Psi^{(l)}_{\mbox{Ai}}(z)
=\begin{pmatrix}
y_{0}(z) & iy_{2}(z) \\
y'_{0}(z) & iy'_{2}(z)
\end{pmatrix},
\end{equation}
and the gauge matrix multiplier $E(z)$ is

\begin{equation}\label{gauge}
E(z) = \sqrt{\pi}
\begin{pmatrix}
\alpha^{-1} & -\alpha \\
-i\alpha^{-1} & -i\alpha
\end{pmatrix}w^{\frac{1}{4}\sigma_{3}}(z).
\end{equation}

We note that, as it follows from (\ref{Phiinf3}) and (\ref{wser}),
the matrix-valued function $E(z)$ is analytic in the disc $B_{d}$.

We are now ready to define an explicit global approximation, $\Phi^{(A)}(z)$, to 
the solution $\Phi^{(1)}(z)$ of the RH problem (1$'''$ - 3$'''$). We take

\begin{equation}\label{PsiAdef}
\Phi^{(A)}(z) =
\left\{\begin{array}{ll}
\Phi^{(\infty)}(z) & \mbox{for} \, \, 
z \notin B_{d}\cup (-B_{d}),\\\\
\Phi^{(z_{0})}(z) & \mbox{for} \, \, 
z \in B_{d},\\\\
\sigma_{3}\Phi^{(z_{0})}(-z)\sigma_{3} & \mbox{for} \, \, 
z \in (-B_{d})
\end{array}\right.
\end{equation}

To see that these formulae indeed provide an approximation to the
solution $\Phi^{(1)}(z)$ we consider the matrix ratio,

\begin{equation}\label{Xdef}
X(z) := \Phi^{(1)}(z)\left(\Phi^{(A)}(z)\right)^{-1}.
\end{equation}
Due to equation (\ref{Phiinfjump}) and the definitions (\ref{Phiend1})
of the parametrix $\Phi^{(z_{0})}(z)$, the function $X(z)$ has no jumps
across the interval $(-z_{0} + d, z_{0} -d)$ and inside the discs
$B_{d}$ and $(-B_{d})$. It is still have jumps across the contour

$$
\Gamma_{0} = (-\infty, -z_{0} - d]\cup (-S_{d})\cup C^{(u)}_{0}
$$

\begin{equation}\label{Gamma0}
\cup \,  C^{(l)}_{0}\cup S_{d}
\cup [z_{0} + d, +\infty),
\end{equation}
where $C^{(u, l)}_{0}$ are the parts of the curves  $C^{(u, l)}$ 
which lie outside of the discs $B_{d}$ and $(-B_{d})$. The curves
$C^{(u, l)}_{0}$ can be taken as straight lines.  The contour
$\Gamma_{0}$ is shown in Figure 3. The matrix-valued function $X(z)$ 
solves the following RH problem posed on the contour $\Gamma_{0}$.

\begin{figure}
\scalebox{0.7}{\includegraphics{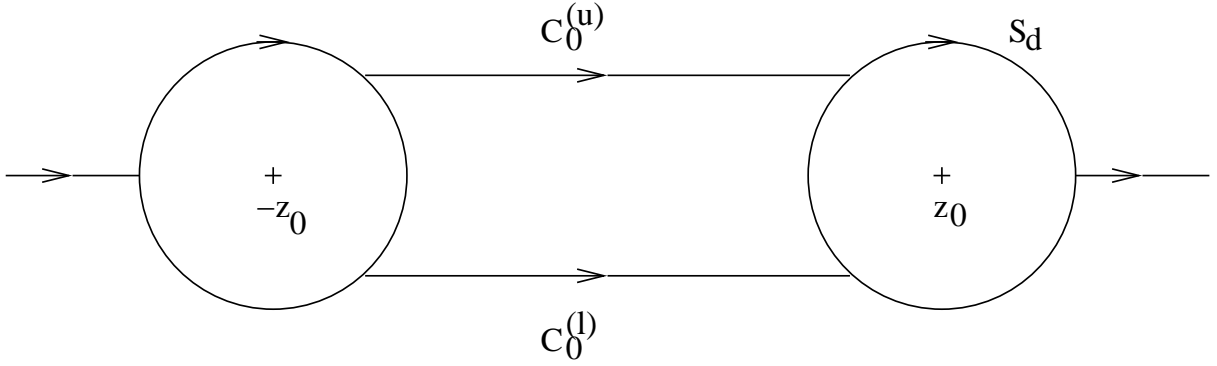}}
\caption{The contour $\Ga_0$.}
\label{fig3}
\end{figure}

\vskip .2in
\noindent
(1$^0$)\quad $X(z)$ is analytic for $z \in {\mathbb C} \setminus 
\Gamma_{0}$, and it has continuous limits, $X_{+}(z)$ and $X_{-}(z)$
from the left and the right of $\Gamma_{0}$.
\vskip .2in
\noindent
(2$^0$)\quad $X(z)$ satisfies the jump condition on $\Gamma_{0}$
\begin{equation}\label{Xjump}
X_{+}(z) = X_{-}(z)G_{X}(z),
\end{equation}
where
\begin{equation}\label{Xjmat}
G_{X}(z) = \left\{\begin{array}{ll}
\Phi^{(\infty)}(z)\begin{pmatrix}
1 & e^{n\left(g_{+}(z) + g_{-}(z) - V_{\lambda} - l\right)} \\
0 & 1
\end{pmatrix}\left(\Phi^{(\infty)}(z)\right)^{-1},& \mbox{for} \, \, 
z \in {\mathbb R}\setminus (-z_{0} - d, z_{0} + d),\\\\
\Phi^{(\infty)}(z)\begin{pmatrix}
1 & 0 \\
e^{-np(z)} & 1
\end{pmatrix}\left(\Phi^{(\infty)}(z)\right)^{-1},
& \mbox{for} \, \, z \in C^{(u)}_{0},\\\\
\Phi^{(\infty)}(z)\begin{pmatrix}
1 & 0 \\
e^{np(z)} & 1
\end{pmatrix}\left(\Phi^{(\infty)}(z)\right)^{-1},
& \mbox{for} \, \, z \in C^{(l)}_{0},\\\\
\Phi^{(z_{0})}(z)
\left(\Phi^{(\infty)}(z)\right)^{-1},
& \mbox{for} \, \, z \in S_{d},\\\\
\sigma_{3}\Phi^{(z_{0})}(-z)
\left(\Phi^{(\infty)}(-z)\right)^{-1}\sigma_{3},
& \mbox{for} \, \, z \in (-S_{d})
\end{array}\right.
\end{equation}
\vskip .2in
\noindent
(3$^0$)\quad as $z \to \infty$, the function $X(z)$ 
has the following uniform asymptotics:
\begin{equation}\label{Xas}
X(z) = I + 0\left(\frac{1}{z}\right),
\quad z \to \infty
\end{equation}
The important feature of this RH problem is that the jump
matrix $G_{X}(z)$ is uniformly close to the identity
matrix as $N \to \infty$. Indeed, using the known asymptotics
of the Airy functions and inequality (\ref{wineq}) one can check directly that the
functions $\Phi^{(z_{0})}(z)$ and $\Phi^{(\infty)}(z)$ match on the
circle $S_{d}$, and the uniform estimate,

\begin{equation}\label{uniform1}
|G_{X}(z) - I | \leq \frac{C}{N}, \quad \mbox{for all}\quad z 
\in  S_{d} \cup (-S_{d}), \quad t \geq 1, \quad \mbox{and}\quad N\geq 1, 
\end{equation}
takes place. Simultaneously, we observe that as $z$ runs over 
${\mathbb R}\setminus (-z_{0} - d, z_{0}
+ d)$, we have 

\begin{equation}\label{uniform2}
0< e^{n\left(g_{+}(z) + g_{-}(z) - V_{\lambda} - l\right)}
\equiv e^{-N\left(
\int_{z_{0}}^{z}(b_{0} + b_{2}s^{2})\sqrt{s^2 - z_{0}^{2}}\,ds\right)}
< e^{-Nc_{0}z^2},
\end{equation}
where the positive constant $c_{0}$ can be choosen the same for all $t\geq 1$
and $N \geq 1$.
Therefore, we conclude that

\begin{equation}\label{uniform3}
|G_{X}(z) - I | \leq Ce^{-Nc_{0}z^2}, \quad \mbox{for all}\quad  z 
\in {\mathbb R}\setminus (-z_{0} - d, z_{0} +d),
\quad t \geq 1, \quad \mbox{and}\quad N\geq 1.
\end{equation}
Finally, inequality (\ref{pderiv}) indicates that on the segments $C_{0}^{(u)}$ and 
$C_{0}^{(l)}$, if they are choosen close enough to the real line, the estimate

\begin{equation}\label{uniform4}
|G_{X}(z) - I | \leq Ce^{-Nc_{0}}, \quad \mbox{for all}\quad  z 
\in  C_{0}^{(u)} \cup C_{0}^{(l)},
\quad t \geq t_{0} > 1, \quad \mbox{and}\quad N\geq 1.
\end{equation}
holds. 

Unlike the estimates (\ref{uniform1}) and (\ref{uniform3}),
estimate (\ref{uniform4}) can not be extended to $t \geq 1$.
However, a slightly weaker version of it is valied for  
$t\geq 1 + 2^{-2/3}N^{-\delta}$ with $\delta < 2/3$. To see this,
let us analyse more carefully  the behavior of the function
Re $p(z)$ near the real line. To this end let us notice that,
in addition to (\ref{pderiv}) we have 

$$
\frac{d^2}{d\sigma^2}\,\mbox{Re}\,p(s+i\sigma)\Big|_{\sigma = 0} =0,
\quad \forall z \in (-z_{0}, z_{0}),
$$
and hence

$$
\mbox{Re}\,p(z) =\sigma \left(
\frac{d}{d\sigma}\,\mbox{Re}\,p(s+i\sigma)\Big|_{\sigma = 0}\right)
+ O(\sigma^{3})
$$
\begin{equation}\label{pser}
= \sigma \left(\frac{2}{\lambda}(b_{0} + b_{2}s^{2})\sqrt{z^2_{0} - s^{2}}\right)
+ O(\sigma^{3}), 
\quad z \equiv s + i\sigma \in  C_{0}^{(u)} \cup C_{0}^{(l)}.
\end{equation}
By a straightforward calculation one can check that

$$
\frac{7}{24}N^{-\delta} \leq b_{0} \leq 1, \quad 
\forall t\geq 1 + 2^{-2/3}N^{-\delta}.
$$
Therefore, equation (\ref{pser}) yields the estimates

\begin{equation}\label{pCu}
n\mbox{Re}\,p(z) \geq c_{0}\sigma N^{1-\delta}
\Bigl( 1 + O\left(\sigma^2 N^{\delta}\right)\Bigr),
\quad z \equiv s + i\sigma \in  C_{0}^{(u)},
\end{equation}
and

\begin{equation}\label{pCl}
n\mbox{Re}\,p(z) \leq c_{0}\sigma N^{1-\delta}
\Bigl( 1 + O\left(\sigma^2 N^{\delta}\right)\Bigr),
\quad z \equiv s + i\sigma \in  C_{0}^{(l)},
\end{equation}
with some positive constant $c_{0}$. 

If we now choose $C^{(u,l)}$ so that

\begin{equation}\label{Ccond}
|\mbox{Im}\,z| \equiv |\sigma| = N^{-1/3}, \quad z \in  C_{0}^{(u)} \cup C_{0}^{(l)},
\end{equation}
and assume 

$$
0 < \delta < \frac{2}{3},
$$
then  (\ref{pCu}) and (\ref{pCl}) 
would imply

\begin{equation}\label{pCu1}
n\mbox{Re}\,p(z) \geq c_{0}N^{2/3-\delta},
\quad z \equiv s + i\sigma \in  C_{0}^{(u)},
\end{equation}
and

\begin{equation}\label{pCl1}
n\mbox{Re}\,p(z) \leq - c_{0} N^{2/3-\delta},
\quad z \equiv s + i\sigma \in  C_{0}^{(l)}.
\end{equation}
(We follow the usual convention to use the same
symbol for perhaps different positive constants whose
exact value is not important to us.) These inequalities in turn
yield the following modification of estimate (\ref{uniform4}).

\begin{equation}\label{uniform5}
|G_{X}(z) - I | \leq Ce^{-c_{0}N^{2/3 - \delta}}, \quad \mbox{for all}\quad  z 
\in  C_{0}^{(u)} \cup C_{0}^{(l)},
\end{equation}
$$
t \geq 1 + 2^{-2/3} N ^{-\delta}, \quad \mbox{and}\quad N\geq 1,
$$
which together with (\ref{uniform1}) and (\ref{uniform3}) lead
to the conclusion that

\begin{equation}\label{jumpest}
||G_{X} - I||_{L^{\infty}(\Gamma_{0})},\quad
||G_{X} - I||_{L^{2}(\Gamma_{0})} \leq \frac{C}{N},
\end{equation}
$$
\mbox{for all} \quad t \geq 1 + 2^{-2/3} N ^{-\delta}, \quad \mbox{and}\quad
N\geq 1
$$
This means that the needed extention of the basic uniform estimate of
the jump matrix has been almost obtained. What is left is the control
of the $t$ - dependence of the estimate. This can be achieved as follows.

Let us attach the subscript ``$\Gauss$" to all the relevant objects,
i.e. the equilibrium measure, the model solutions, etc., which correspond
to the gaussian potential, $V_{\Gauss}(z) = z^2$. By the very nature 
of our approach, as $t \to \infty$, all the main ingredients of the above
scheme, i.e.

$$
g(z),\quad \Phi^{(\infty)}(z), \quad  \Phi^{(z_{0})}(z),
 $$
converge to the respective $\Gauss$ - quantities, i.e. to

$$
g_{\Gauss}(z),\quad, 
\Phi_{\Gauss}^{(\infty)}(z) \quad  \Phi_{\Gauss}^{(z_{0})}(z).
$$
Moreover, the following inequalities for the jump matrix of
the $X$ - RH problem  can be established by a straightforward
calculations.

\begin{equation}\label{gauss1}
| G_{X}(z)G^{-1}_{X_{\Gauss}}(z) - I|
\leq \frac{CNe^{-c_{0}Nz^2}z^{4}}{t}, \quad \mbox{for all}\quad z \in 
{\mathbb R}\setminus (-z_{0} - d, x_{0} + d),
\end{equation}
$$
t \geq 1 + 2^{-2/3}N^{-\delta}, \quad \mbox{and}\quad N \geq 1,
$$

\begin{equation}\label{gauss2}
|G_{X}(z)G^{-1}_{X_{\Gauss}}(z) - I|
\leq \frac{CNe^{-c_{0}N^{2/3 -\delta}}}{t}, \quad \mbox{for all}\quad z \in 
C^{(u)}_{0}\cup C^{(l)}_{0}
\end{equation}
$$
t \geq 1 + 2^{-2/3}N^{-\delta}, \quad \mbox{and}\quad N \geq 1,
$$

\begin{equation}\label{gauss3}
|G_{X}(z)G^{-1}_{X_{\Gauss}}(z) - I|
\leq \frac{C}{tN}, \quad \mbox{for all}\quad z \in 
S_{d}\cup (-S_{d}),
\end{equation}
$$
t \geq 1 + 2^{-2/3}N^{-\delta}, \quad \mbox{and}\quad N \geq 1,
$$
Put

\begin{equation}\label{Xtidef}
\tilde{X}(z) = X(z)X^{-1}_{\Gauss}(z).
\end{equation}
The function $\tilde{X}(z)$ solves the RH problem on the same contour
$\Gamma_{0}$ as the function $X(z)$ and with the jump matrix,

$$
G_{\tilde{X}}(z) \equiv G_{X}(z)G^{-1}_{X_{\Gauss}}(z).
$$
The inequalities (\ref{gauss1} - \ref{gauss3}) yield then the
following modification of estimate (\ref{jumpest}),

\begin{equation}\label{gauss4}
\|G_{\tilde{X}} - I\|_{L^{\infty}(\Gamma_{0})},\quad
\|G_{\tilde{X}} - I\|_{L^{2}(\Gamma_{0})} \leq \frac{C}{tN},
\end{equation}
$$
\mbox{for all} \quad t \geq 1 + 2^{-2/3} N ^{-\delta}, \quad \mbox{and}\quad
N\geq 1
$$

The proof of Proposition \ref{onecut} can be now completed in the usual
way, by iterating the $\tilde{X}$ - RH problem(cf. \cite{DKMVZ} and \cite{EM}).
Indeed, by iterative arguments, we can see that for any $K \geq 0$,

\begin{equation}\label{gauss5}
\left|\tilde{X}(z) - I -\sum_{k=1}^K N^{-k}
f_k\left(\frac{n}{N};t, z\right)\right|\le \frac{C(K)}{tN^{K+1}(1+|z|)}, 
\end{equation}
and also

\begin{equation}\label{gauss55}
\left|f_k\left(\frac{n}{N};t, z\right)\right|\le \frac{C(k)}{t(1+|z|)}.
\end{equation} 
Denote $m_{1}^{\infty}$, $m_{1,\Gauss}$ and $\tilde{m}_{1}$ the matrix coefficients
of the terms $1/z$ in the asymptotic series at $z = \infty$ of the
functions $\Phi^{(\infty)}(z)$, $X_{\Gauss}(z)$ and $\tilde{X}(z)$,
respectively. Then, for the coefficient $m_{1}^{(n)}$ of series (\ref{Yasymp})
we will have from (\ref{first}), (\ref{Xdef}), and (\ref{Xtidef}) that

\begin{equation}\label{gauss6}
e^{-\frac{nl}{2}\sigma_{3}}m_{1}^{(n)}e^{\frac{nl}{2}\sigma_{3}}
= m_{1}^{\infty} + m_{1,\Gauss} + \tilde{m}_{1}.
\end{equation}
In virtue of estimates (\ref{gauss5}) and (\ref{gauss55}) we have that

\begin{equation}\label{gauss555}
\left|\tilde{m}_{1} - I -\sum_{k=1}^K N^{-k}
r_k\left(\frac{n}{N};t\right)\right|\le C(K)N^{-K-1}t^{-1}, 
\end{equation}

\begin{equation}\label{gauss5555}
\left|r_k\left(\frac{n}{N};t\right)\right|\le C(k)t^{-1},
\end{equation} 
while

\begin{equation}\label{gauss7}
m_{1,\Gauss} = 
e^{-\frac{nl_{\Gauss}}{2}\sigma_{3}}m_{1, \Gauss}^{(n)}
e^{\frac{nl_{\Gauss}}{2}\sigma_{3}} - m_{1, \Gauss}^{\infty}.
\end{equation}
Observe now that the matrices $m_{1}^{\infty}$, $m_{1, \Gauss}^{\infty}$, and
$m_{1,\Gauss}^{(n)}$  can be  evaluated explicitly. Indeed,
the first two can be obtained from (\ref{Phiinf2}), taking into account
that $z_{0,\Gauss} = \sqrt{2\lambda}$, and the third one follows
from the fact that the normalizing constants $h_{n, \Gauss}$ are known -
see (\ref{fin7}).  Therefore, performing the calculations
indicated, we derive from equations (\ref{gauss555}), (\ref{gauss6}),
and (\ref{coefY}) the following estimates for the recurrence coefficients $R_{n}$,

\begin{equation}\label{gauss8}
\left|R_{n}(t) - \frac{z_{0}^{2}}{4} -\sum_{k=1}^K N^{-k}
f_k\left(\frac{n}{N};t\right)\right|\le C(K)N^{-K-1}t^{-1}, 
\end{equation}

\begin{equation}\label{gauss9}
\left|f_k\left(\frac{n}{N};t\right)\right|\le C(k)t^{-1}.
\end{equation} 

$$
\mbox{for all}\quad t \geq 1 + 2^{-2/3}N^{-\delta},\quad n = N-1, N, N+1, 
\quad \mbox{and}\quad N \geq 1,
$$

Finally, repeating the arguments we used in the proof of Theorem \ref{asympt},
we conclude that the odd coefficients $f_{k}$ in the series from (\ref{gauss8})
are actually absent. The Proposition \ref{onecut} follows.

\end{document}